\documentclass[journal=jacsat,manuscript=article]{achemso}

\usepackage[version=3]{mhchem} 
\usepackage{graphicx} %
\usepackage{float} %
\usepackage{subfigure} %
\usepackage{svg}
\usepackage{multirow}
\usepackage{multicol}
\usepackage{booktabs}
\usepackage[utf8]{inputenc}
\usepackage{diagbox}
\usepackage{textcomp} 
\usepackage{gensymb}  
\usepackage{longtable}
\usepackage{caption}
\usepackage{subcaption}
\usepackage{multirow}
\usepackage{xr}
\newcommand*{\citen}[1]{%
  \begingroup
    \romannumeral-`\x 
    \setcitestyle{numbers}%
    \cite{#1}%
  \endgroup   
}
\author{Ziwei Chai}
\email{ziwei.chai@chem.uzh.ch}
\author{Sandra Luber}
\affiliation[University of Zurich]
{Department of Chemistry, University of Zurich, Winterthurerstrasse 190, 8057 Zurich, Switzerland}

\title[An \textsf{achemso} demo]
  {The anisotropic interface continuum solvation model and the finite-element anisotropic Poisson solver.} 

\abbreviations{GCE}

\begin{document}

\quad

\begin{abstract}
We propose an anisotropic interfacial continuum solvation (AICS) model to simulate the distinct in-plane and out-of-plane dielectric constants of liquids near solid–liquid interfaces and their spatial variations along the surface normal direction. In low-electron-density regions, each dielectric function in the diagonal components of a dielectric tensor varies monotonically with distance from the solid surface along the surface normal direction; in high-electron-density regions near the surface, each dielectric function adopts the electron-density-based formulation proposed by Andreussi et al. (J. Chem. Phys. 136, 064102 (2012)) The resulting dielectric tensor is continuously differentiable with respect to both electron density and spatial coordinates. We derived analytical expressions for electrostatic contributions to the Kohn–Sham potential and atomic forces, and implemented the AICS model, including these analytical derivatives, into the CP2K software package. To solve the anisotropic Poisson equations, we developed a parallel finite-element anisotropic Poisson solver (FEAPS) based on the FEniCSx platform and its interface with CP2K. Analytical forces were validated against finite-difference calculations, while electrostatic potentials computed under vacuum and isotropic solvent conditions using AICS and FEAPS were benchmarked against standard vacuum DFT and SCCS results, respectively. In the anisotropic solvent environment characterized by the enhanced in-plane and reduced out-of-plane dielectric functions near the Ag(111) surface, we calculated the resulting work functions and electrostatic potentials, and optimized the adsorption geometry for OH*. Compared to the isotropic case, we observed more pronounced work function shifts and spatially modulated electrostatic profiles across different charge states. Our results also showed that OH* tilted more towards the plane parallel to the surface under the anisotropic dielectric conditions.
\end{abstract}

\section{1. Introduction}
The dielectric behavior of a liquid near a solid surface can differ significantly from that in the bulk phase. For example, while the dielectric constant of bulk water is isotropic and spatially homogeneous, experimental\cite{science_low_dielec_water_2018, naturecomm_negative_diele_2019, in_plane_arxiv_2024} and simulation\cite{anisotropic_profile_jcp_2003, polar_anisotropic_cal_jcp_2005, diamond_anisotropic_profile_prl_2011, polarball_anisotropic_profile_prl_2016, md_spatial_nanoconfined_jcp_2016, jpcb_2020, pre_confined_water_2020, nanoscale_negative_2021, jpcb_nacl_confined_water_2021, jpcl_low_z_dielectric_2021, G_monet_prl_2021, recent_anisotropic_dielectric_PRL_2023, jiaxin_Kathirina_2025} studies indicated that the dielectric functions of interfacial water exhibit pronounced anisotropy and vary spatially along the surface normal direction. It has been commonly shown in the studies of nanoconfined water between two surfaces that the in-plane dielectric constant is higher than that of bulk water\cite{in_plane_arxiv_2024, polar_anisotropic_cal_jcp_2005, diamond_anisotropic_profile_prl_2011, polarball_anisotropic_profile_prl_2016, jpcb_2020, nacl_solution_jpcb_md_tensor_2024}, while the out-of-plane dielectric constant is lower\cite{science_low_dielec_water_2018, polar_anisotropic_cal_jcp_2005, diamond_anisotropic_profile_prl_2011, polarball_anisotropic_profile_prl_2016, jpcb_2020, pre_confined_water_2020, jpcb_nacl_confined_water_2021, jpcl_low_z_dielectric_2021, G_monet_prl_2021,  nacl_solution_jpcb_md_tensor_2024} or even negative\cite{naturecomm_negative_diele_2019, polar_anisotropic_cal_jcp_2005, diamond_anisotropic_profile_prl_2011, polarball_anisotropic_profile_prl_2016, nanoscale_negative_2021, jpcl_low_z_dielectric_2021}, where the dielectric response incorporates contributions from bond stretching and torsion, as well as molecular rotation. Tran et al. recently developed a classical molecular dynamics framework that efficiently captured metal polarization and incorporated a novel ac field method to probe the local dielectric response of interfacial water at the frequency of 2 GHz\cite{recent_anisotropic_dielectric_PRL_2023}. Their simulation results explicitly demonstrated for the first time that, at the metal-water interface, the in-plane dielectric constant of interfacial water, within a few atomic layers, exceeds that of bulk water, while the out-of-plane one is lower or even negative. Additionally, the dielectric constants vary with distance from the metal surface. Their computed average out‑of‑plane dielectric constant agreed with reported measurements on confined thin films.\\
The implicit solvent method is a fast and effective simulation technique for simulating solvents at solid-liquid interfaces, representing the surrounding liquid solvent as a continuous, polarizable medium\cite{implicit_solvation_review_stefan_karsten}. It implicitly integrates the degrees of freedom of the solvent or dissolved ions, eliminating the need for explicit sampling and greatly reducing computational cost\cite{implicit_solvation_review_stefan_karsten, solvent_aware_jctc_2019}. Efforts have been made in the quantum chemistry community to incorporate anisotropic solvation effects into the polarizable continuum model (PCM) by replacing the scalar dielectric function with a dielectric tensor\cite{old_ten_1_jpcb_1997, old_ten_2_jcp_1997, old_ten_3_1998, old_ten_4_2003, jdftx_tensor_input}. In addition, the spatial variation of the dielectric function around a molecule has been directly modeled, enabling more accurate simulations of molecular or ionic solvation at interfaces, such as the air–water boundary\cite{pcm_1, pcm_2, pcm_3, pcm_4}. However, to the best of our knowledge, apart from a study that considered the anisotropy of the solvent dielectric constant in a continuum model near solid surfaces using a pre‑parameterized, non‑self‑consistent dielectric function and a one‑dimensional extended Poisson equation\cite{oxford_aniso_poisson_jpcb_2013}, nearly all widely used implicit solvent methods and their implementations in DFT packages for periodic solid-liquid interface simulations neglected this anisotropy, instead treating the solvent dielectric tensor as a scalar field\cite{sccs_jcp_2012, vasp_jdft_jcp_2014, jdft_jcpb_2005, solvent_aware_jctc_2019, soft_sphere_qe_jctc_2017, bigdft_jpcc_2020, gpaw_periodic_prb_2006, gpaw_cavity_jcp_2014, gpaw_gce_jcp_2018, fattebert_gygi_ijqc_2003, pwmat_jc_2020, jdftx_2017, candle_jdftx_jcp_2015, wenjin_2017, onetep_is_jctc_2018, onetep_is_el_2011, fhiaims_mpb_jctc_2016, fhiaims_asc_jctc_2017, crystal_is_jctc_2018}. Furthermore, these models did not account for the variation of the dielectric tensor with distance from the solid surface.\\
In light of the unique solvent characteristics at solid-liquid interfaces and the limitations of existing models in capturing both dielectric anisotropy and spatial variation\cite{nacl_solution_jpcb_md_tensor_2024}, a more advanced implicit solvation method is urgently needed. Here, we propose the anisotropic interface continuum solvation (AICS) method for simulating solid–liquid interfaces. Unlike traditional models that treat the solvent’s dielectric function as a scalar field, our approach represents it as a tensor field to account for dielectric anisotropy and additionally allows for spatial variation along the surface normal to more realistically mimic the permittivity distribution. This tensor is continuously differentiable with respect to both the quantum mechanical degrees of freedom and the spatial coordinates along the surface normal, enabling precise computation of analytical atomic forces and the Kohn–Sham potential. To solve the anisotropic generalized Poisson equation arising from this formulation, we developed a finite element anisotropic Poisson solver (FEAPS)—an MPI-parallel solver based on FEniCSx/DOLFINx\cite{BarattaEtal2023, ScroggsEtal2022, BasixJoss, AlnaesEtal2014}—which ensures both accuracy and efficiency in handling anisotropic electrostatics. The AICS method, FEAPS, and its coupling interface have been implemented and integrated into the open-source software package CP2K\cite{cp2k_quickstep_cpc_2005}.\\
The remainder of this paper is organized as follows. Sec. 2 presents the theoretical formalism of anisotropic Poisson equation, electron-density- and surface-distance-dependent dielectric function near the solid surface, and their contributions to Kohn-Sham potential and analytical atomic forces, for the GPW method in the CP2K software package. Sec. 3 presents the algorithm and program details of the finite element anisotropic Poisson solver based on Python and Dolfinx (FEniCSx), and its interface with the CP2K software package. Sec. 4 presents the validation and test results, including: analytical force accuracy (4.1); electrostatic potentials of the Ag(111) surface in vacuum (4.2.1), in isotropic dielectric water (4.2.2), and in anisotropic dielectric water (4.2.3); OH* adsorption on Ag(111) in anisotropic dielectric water (4.3); and computational performance and scaling (4.4). We summarize and conclude our work in Sec. 5.\\
\section{2. Anisotropic and Surface-Distance-Dependent Dielectric Function: Formulation and Derivatives}
\subsection{2.1 Generalized Poisson Equation with a Dielectric Tensor}
For an anisotropic dielectric continuum, the permittivity is represented as a tensor field rather than a scalar field. When the solute's charge density $\rho^{\text {solute}}(\boldsymbol{r})$ is embedded in such a solvent medium, the Poisson equation for the total electrostatic potential $\phi^{\text {tot}}(\boldsymbol{r})$ of the explicit solute/implicit solvent system is given by\cite{old_ten_3_1998}
\begin{equation}
\nabla \cdot\left(\boldsymbol{\epsilon}(\boldsymbol{r}) \nabla \phi^{\text {tot}}(\boldsymbol{r})\right)=-4 \pi \rho^{\text {solute}}(\boldsymbol{r}).
\label{eq1}
\end{equation}
$\boldsymbol{\epsilon}(\boldsymbol{r})$ is a dielectric matrix field that transforms (or acts on) the gradient of the electrostatic potential $\nabla \phi^{\text {tot}}(\boldsymbol{r})$, which takes the general form of\cite{old_ten_3_1998}
\begin{equation}
\boldsymbol{\epsilon}(\boldsymbol{r})=\left(\begin{array}{ccc}
\epsilon_{\text {xx}}(\boldsymbol{r}) & \epsilon_{\text {xy}}(\boldsymbol{r}) & \epsilon_{\text {xz}}(\boldsymbol{r}) \\
\epsilon_{\text {yx}}(\boldsymbol{r}) & \epsilon_{\text {yy}}(\boldsymbol{r}) & \epsilon_{\text {yz}}(\boldsymbol{r}) \\
\epsilon_{\text {zx}}(\boldsymbol{r}) & \epsilon_{\text {zy}}(\boldsymbol{r}) & \epsilon_{\text {zz}}(\boldsymbol{r})
\end{array}\right).
\label{eq2}
\end{equation}
x and y represent two non-parallel coordinate basis vectors parallel to the solid surface, while z represents the coordinate basis vector perpendicular to the solid surface. Each diagonal element $\epsilon_{\text {ii}}(\boldsymbol{r})(\text {i} \in\{\text {x}, \text {y}, \text {z}\})$ represents the dielectric response of the continuum along the i-direction at position $\boldsymbol{r}$ when an electric field is applied in the same direction. Each off-diagonal element $\epsilon_{\text {ij}}(\boldsymbol{r})(\text {i}, \text {j} \in\{\text {x}, \text {y}, \text {z}\}, \text {i} \neq \text {j})$ represents the dielectric response of the dielectric continuum in the direction i when an electric field is applied in the direction $\text {j}$. In this work, we disregarded the contributions from all off-diagonal elements $\epsilon_{\text {ij}}(\boldsymbol{r})(\text {i}, \text {j} \in\{\text {x}, \text {y}, \text {z}\}, \text {i} \neq \text {j})$ by setting them to zero. This implies that we neglect cross-coupling: it is assumed that an electric field along the j-direction does not produce an appreciable dielectric response along the i-direction.\\
The electrostatic energy $E^\text {H}$ of the explicit solute/implicit solvent system expressed in terms of the electric field $\boldsymbol{E}(\boldsymbol{r})$ and the electric displacement field $\boldsymbol{D}(\boldsymbol{r})$ can be derived as follows\cite{sccs_jcp_2012, implicit_solvation_review_stefan_karsten, jackson1998classical}
\begin{equation}
\begin{aligned}
E^\text {H}&=\frac{1}{8 \pi} \int \boldsymbol{E}(\boldsymbol{r}) \cdot \boldsymbol{D}(\boldsymbol{r}) d \boldsymbol{r}\\
&=\frac{1}{8 \pi} \int \nabla \phi^{\text {tot}}(\boldsymbol{r}) \cdot\left(\boldsymbol{\epsilon}(\boldsymbol{r}) \nabla \phi^{\text{tot}}(\boldsymbol{r})\right) d \boldsymbol{r}\\
&=-\frac{1}{8 \pi} \int \phi^{\text {tot}}(\boldsymbol{r}) \nabla \cdot\left(\boldsymbol{\epsilon}(\boldsymbol{r}) \nabla \phi^{\text{tot}}(\boldsymbol{r})\right) d \boldsymbol{r}.
\label{eq3}
\end{aligned}
\end{equation}
$\phi^{\text{tot}}(\boldsymbol{r})$ is the electrostatic potential. The derivation of the second equality in Eq.~\eqref{eq3} comes from the product rule and neglecting the surface term $\int \nabla \cdot\left(\phi^{\text{tot}}(\boldsymbol{r}) \boldsymbol{\epsilon}(\boldsymbol{r}) \nabla \phi^{\text{tot}}(\boldsymbol{r})\right) d \boldsymbol{r}$ \cite{implicit_solvation_review_stefan_karsten, my_cpc_2025}. According to Eq.~\eqref{eq1}, Eq.~\eqref{eq3} can be written as
\begin{equation}
E^\text {H}=\frac{1}{2} \int \phi^{\text{tot}}(\boldsymbol{r}) \rho^{\text {solute}}(\boldsymbol{r}) d \boldsymbol{r},
\label{eq4}
\end{equation}
where $\rho^{\text {solute}}(\boldsymbol{r})$ is the sum of the electron density $\rho^{\text {el}}(\boldsymbol{r})$ and the charge density $\rho^{\text {ion}}(\boldsymbol{r})$ of the atomic nuclei and the core electrons.\\
\subsection{2.2 Electron-Density- and Surface-Distance-Dependent Dielectric Function}
\subsubsection{2.2.1 The Generalization of Andreussi et al.’s Electron-Density-Dependent Dielectric Function}
\begin{figure}[H]
\centering
\includegraphics[width=\textwidth]{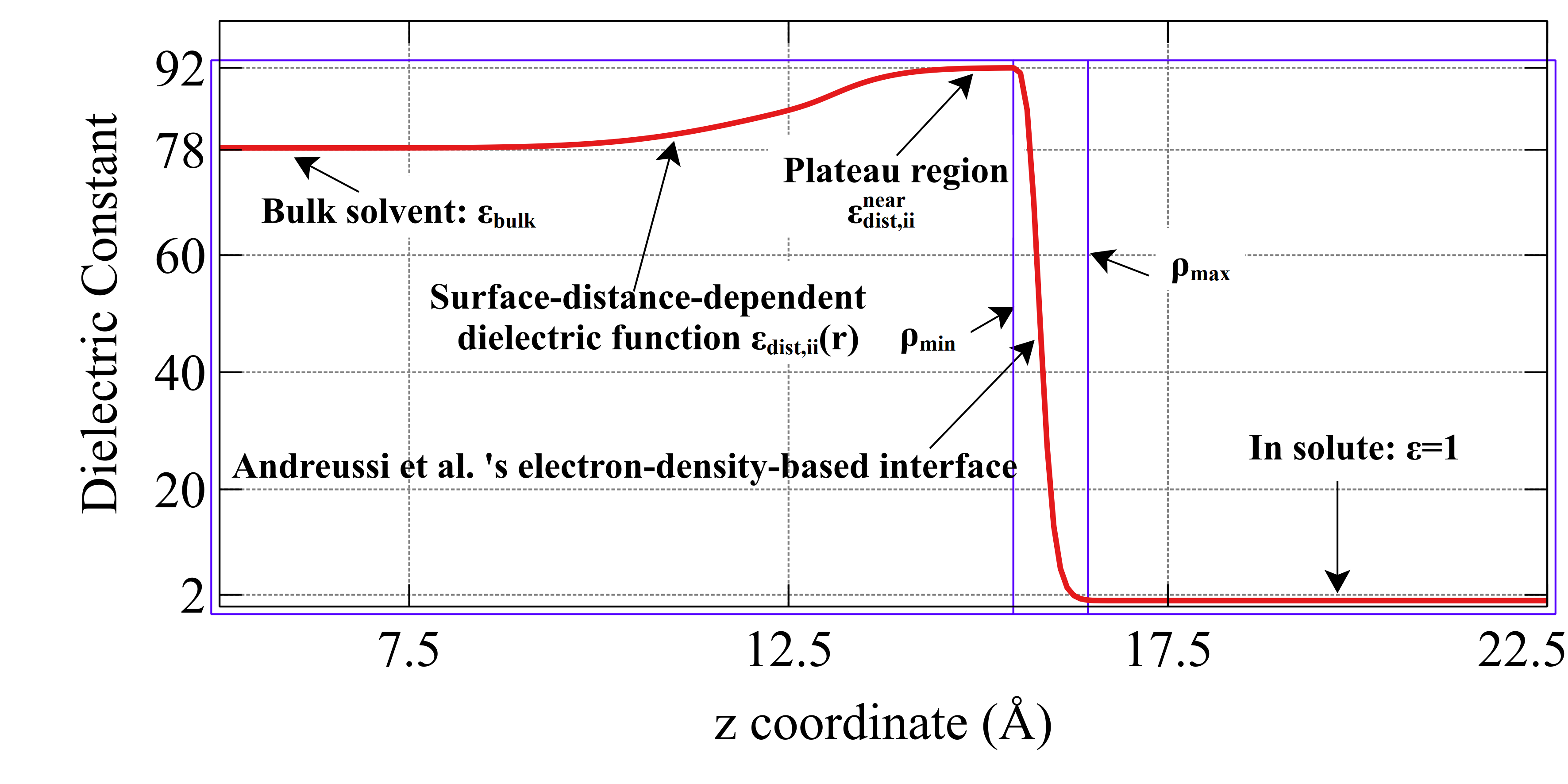}
\caption{Schematic illustration of an exemplary dielectric function $\epsilon_{\text {ii}}(\boldsymbol{r})$ formulated in this work. The red curve represents the in-plane planar-averaged values of a representative in-plane dielectric function ($\epsilon_{\text {xx}}(\boldsymbol{r})$ or $\epsilon_{\text {yy}}(\boldsymbol{r})$) as a function of the z coordinate.}\label{fig0.0}
\end{figure}

Building upon the electron-density-dependent dielectric formulation proposed by Andreussi et al. in the self-consistent continuum solvation (SCCS) model\cite{sccs_jcp_2012}, we extend the treatment of the solvent dielectric function to incorporate both spatial variation in the bulk solvent region and tensorial anisotropy allowing distinct $\epsilon_{\text {xx}}(\boldsymbol{r})$, $\epsilon_{\text {yy}}(\boldsymbol{r})$, and $\epsilon_{\text {zz}}(\boldsymbol{r})$. In the original SCCS approach, the dielectric function varies smoothly as a function of electron density from 1 (vacuum dielectric constant) in the high-electron-density region to the dielectric constant of the bulk solvent in the low-electron-density region. In this work, we retain this electron-density-based interpolation in the high-electron-density region, while allowing the dielectric tensor in the low-electron-density region to vary smoothly and anisotropically along the surface normal direction, in order to more realistically capture interfacial solvent behavior.\\
For each diagonal element $\epsilon_{\text {ii}}(\boldsymbol{r})(\text {i} \in\{\text {x,y,z}\})$ of the dielectric tensor in Eq.~\eqref{eq2}, when the $\rho^{\text {el}}(\boldsymbol{r})$ at $\boldsymbol{r}$ is greater than $\rho_{\max}$, $\epsilon_{\text {ii}}(\boldsymbol{r})$ is set to 1 (as in Ref. \citen{sccs_jcp_2012}), whereas when the $\rho^{\text {el}}(\boldsymbol{r})$ at $\boldsymbol{r}$ is less than $\rho_{\min}$, $\epsilon_{\text {ii}}(\boldsymbol{r})$ is set to the value of the function $\epsilon_{\text {dist,ii}}(\boldsymbol{r})$ at $\boldsymbol{r}$, as illustrated in Fig.~\ref{fig0.0}. $\rho_{\max}$ and $\rho_{\min}$ are maximal and minimal electron density threshold values, respectively. In the region where the electron density falls below $\rho_{\min}$, the surface-distance-dependent dielectric function $\epsilon_{\text{dist,ii}}(\boldsymbol{r})$ is nearly constant in the vicinity of the $\rho_{\min}$ boundary, approaching a fixed value denoted as $\epsilon_{\text{dist,ii}}^{\text{near}}$. When the electron density $\rho^{\text {el}}(\boldsymbol{r})$ lies between $\rho_{\max}$ and $\rho_{\min}$, the formulation proposed by Andreussi et al. in the SCCS method is used, leading to the final expression, for $\epsilon_{\text {dist,ii}}^{\text {near}}>0$\cite{sccs_jcp_2012}
\begin{equation}
\epsilon_{\text {ii}}(\boldsymbol{r})=\left\{\begin{array}{cc}
1 & \rho^{\text {el}}(\boldsymbol{r})>\rho_{\max} \\
e^{t_{\text {ii}}\left(\rho^{\text {el}}(\boldsymbol{r})\right)} & \rho_{\max}\geq\rho^{\text {el}}(\boldsymbol{r})\geq\rho_{\min} \\
\epsilon_{\text {dist,ii}}(\boldsymbol{r}) & \rho_{\min}>\rho^{\text {el}}(\boldsymbol{r})
\end{array}\right.,
\label{eq5}
\end{equation}
in which the form of function $t_{\text {ii}}$ is the one defined in Eq. 42 in Ref. \citen{sccs_jcp_2012}, except for that the dielectric function of bulk solvent was replaced by $\epsilon_{\text {dist,ii}}^{\text {near}}$
\begin{equation}
t_{\text {ii}}\left(\rho^{\text {el}}(\boldsymbol{r})\right)=\frac{\ln \epsilon_{\text {dist,ii}}^{\text {near}}}{2 \pi}\left[2 \pi \frac{\left(\ln \rho_{\text {max}}-\ln \rho^{\text {el}}(\boldsymbol{r})\right)}{\left(\ln \rho_{\text {max}}-\ln \rho_{\text {min}}\right)}-\sin \left(2 \pi \frac{\left(\ln \rho_{\text {max}}-\ln \rho^{\text {el}}(\boldsymbol{r})\right)}{\left(\ln \rho_{\text {max}}-\ln \rho_{\text {min}}\right)}\right)\right] .
\label{eq6}
\end{equation}
Clearly, $\epsilon_{\text {ii}}$ is continuously differentiable with respect to $\rho^{\text {el}}$ which is continuously differentiable with respect to the coordinates. Furthermore, if $\epsilon_{\text{dist,ii}}(\boldsymbol{r})$ is continuously differentiable in space, then so is $\epsilon_{\text{ii}}(\boldsymbol{r})$. To retain formal generality and allow for the possibility of negative dielectric function values, we define $\epsilon_{\text {ii}}(\boldsymbol{r})$ and $t_{\text {ii}}$ by Eqs.~\ref{eq7} and \ref{eq8} for $\epsilon_{\text{dist,ii}}^{\text{near}}<0$. In this work, however, we do not consider scenarios where the dielectric function becomes negative. The forms of Eqs.~\eqref{eq5}--\eqref{eq6} and Eqs.~\eqref{eq7}--\eqref{eq8} ensure that, as the electron density of the solute decreases, the dielectric function rapidly transitions from the vacuum dielectric constant (1) to $\epsilon_{\text {dist,ii}}^{\text {near}}$, as illustrated in Fig.~\ref{fig0.0}.\\
\subsubsection{2.2.2 The Replacement of Solvent’s Bulk Dielectric Constant to a Surface-Distance-Dependent Dielectric Function}
The dielectric constant of the solvent near solid surfaces can vary notably over distances of at least several nanometers along the surface normal direction\cite{recent_anisotropic_dielectric_PRL_2023, jiaxin_Kathirina_2025}. As a representative case, water was shown in recent simulations to exhibit strong dielectric anisotropy near solid surfaces: the in-plane components of the dielectric tensor ($\epsilon_{\text{xx}}$ and $\epsilon_{\text{yy}}$) increase to around 90, while the out-of-plane component ($\epsilon_{\text{zz}}$) decreases to around -30, reflecting the combined effects of rotational and vibrational modes, accounting for contributions from both molecular rotations and bond vibrations.\cite{recent_anisotropic_dielectric_PRL_2023}. Similar characters of the dielectric constants of water near surface were also indicated in many experimental and computational studies for nanoconfined water, as mentioned in Sec. 1. Therefore, a natural and desirable requirement is that the solvent dielectric tensor component $\epsilon_{\text{ii}}(\boldsymbol{r})$ in this region is not simply a bulk solvent dielectric constant but rather varies monotonically—either increasing or decreasing—with the distance from the solid surface to realistically mimic the distribution. In the plane parallel to the metal surface, $\epsilon_{\text{ii}}(\boldsymbol{r})$ can take the same value.\\
Due to the nature of the exponential decay of electron density away from the solid surface, the generalization of Andreussi et al.’s electron-density-dependent dielectric function\cite{sccs_jcp_2012} is suitable for representing the narrow transition region between the inner solute region and low-electron-density region. To preserve the desirable feature of the SCCS method, ensuring that the dielectric function remains continuously differentiable with respect to both the quantum mechanical degrees of freedom and spatial coordinates, the derivative of $\epsilon_{\text {dist,ii}}(\boldsymbol{r})$ along the surface normal direction should be (quite close to) zero where $\rho^{\text {el}}(\boldsymbol{r})$ is equal to $\rho_{\min}$. Additionally, $\epsilon_{\text {dist,ii}}(\boldsymbol{r})$ can be independent of the electron density and depend solely on the distance from the outermost solid surface atomic layer, while remaining continuously differentiable with respect to spatial coordinates. In addition to satisfying the above conditions, the parameters of $\epsilon_{\text {dist,ii}}(\boldsymbol{r})$ should ensure a certain degree of flexibility in its distribution along the surface normal direction.\\
Based on the above considerations, we have designed the following monotonical functional form for $\epsilon_{\text {dist,ii}}(\boldsymbol{r})$:
\begin{equation}
\epsilon_{\text {dist,ii}}(\boldsymbol{r})=\epsilon_{\text {dist,ii}}\left(r_\text {z}\right)=\epsilon_{\text {dist,ii}}^{\text {near}}+0.5 \times\left(\epsilon_{\text {bulk}}-\epsilon_{\text {dist,ii}}^{\text {near}}\right) \times\left(1+\operatorname{erf}\left(\frac{d-d_\text {j}^\text {0}}{\sigma}\right)\right),
\label{eq9}
\end{equation}
with respect to $d$. $d$ is the rescaled distance of $\boldsymbol{r}$ from the solid surface along z-direction after undergoing nonlinear transformation, if the z-coordinate in the solvent region is greater than that in the bulk solute region,
\begin{equation}
d=r_\text {z}-r_\text {z}^\text {ref}+\frac{\left(r_\text {z}-r_\text {z}^\text {ref}-d_\text {i}^\text {0}\right)}{1+\alpha  \tanh \left(\beta \left(r_\text {z}-r_\text {z}^\text {ref}-d_\text {i}^\text {0}\right)\right)},
\label{eq10}
\end{equation}
and if the z-coordinate in the solvent region is smaller than that in the solid solute region,
\begin{equation}
d=r_\text {z}^\text {ref}-r_\text {z}+\frac{\left(r_\text {z}^\text {ref}-r_\text {z}-d_\text {i}^\text {0}\right)}{1+\alpha  \tanh \left(\beta \left(r_\text {z}^\text {ref}-r_\text {z}-d_\text {i}^\text {0}\right)\right)}.
\label{eq11}
\end{equation}
$r_\text{z}$ is the z-component of $r$, and $r_\text{z}^{\text{ref}}$ serves as a reference for determining distances. It should be noted that $d$ only depends on $r_\text{z}$ and is independent of $r_\text{x}$ and $r_\text{y}$, so $d$ is uniform in the xy-plane. $r_\text{z}^{\text{ref}}$ can be chosen, for example, as the averaged z-coordinates of the atoms in the outermost atomic layer of the solid surface. $d_\text {i}^\text {0}$ is the distance $\left|r_\text {z}-r_\text {z}^\text {ref}\right|$ at which $\epsilon_{\text {dist,ii}}(\boldsymbol{r})$ equals $\frac{\epsilon_{\text{bulk}}+\epsilon_{\text {dist,ii}}^{\text{near}}}{2}$. $\epsilon_{\text {bulk}}$ is the dielectric function of the bulk solvent. $\sigma$ controls the width of the transition region of the curve between $\epsilon_{\text {dist,ii}}^{\text {near}}$ and $\epsilon_{\text {bulk}}$, and a larger $\sigma$ value results in a wider transition region. $\alpha$ and $\beta$ control the strength of the nonlinear transformation of $\left|r_\text {z}-r_\text {z}^\text {ref}\right|$. $\alpha$ usually should be a value between -1 and 1. If $\beta$ is positive, a larger positive (small negative) $\alpha$ will result in a sharper (smoother) transition region between $\epsilon_{\text {dist,ii}}^{\text {near}}$ and $\frac{\epsilon_{\text{bulk}}+\epsilon_{\text {dist,ii}}^{\text{near}}}{2}$, and a smoother (sharper) transition between $\frac{\epsilon_{\text{bulk}}+\epsilon_{\text {dist,ii}}^{\text{near}}}{2}$ and  $\epsilon_{\text {bulk}}$. If $\alpha$ is positive, changing $\beta$ has the similar effect to the sharpness of the transition region. Fig.~\ref{fig0.0} illustrates the variation of $\epsilon_{\text {ii}}(\boldsymbol{r})$ as a function of the distance from the surface, and Fig.~\ref{figs0} shows the shapes of the functions under different parameter settings.\\
\subsubsection{2.2.3 Derivation of Electrostatic Contributions to the Kohn–Sham Potential and Forces}
From the definition of a functional derivative,
\begin{equation}
\lim _{\lambda \rightarrow 0} \frac{E_{\left[\rho^{\text {el}}(\boldsymbol{r})+\lambda f(\boldsymbol{r})\right]}^\text{H}-E_{\left[\rho^{\text {el}}(\boldsymbol{r})\right]}^\text{H}}{\lambda}=\int \frac{\delta E^\text {H}}{\delta \rho^{\text {el}}(\boldsymbol{r})} f(\boldsymbol{r}) d \boldsymbol{r},
\label{eq12}
\end{equation}
one can derive the functional derivative $\frac{\delta E^\text {H}}{\delta \rho^{\text {el}}(\boldsymbol{r})}$ by explicitly deriving the limit term. $E_{\left[\rho^{\text {el}}(\boldsymbol{r})\right]}^\text{H}$ here formally means that the electrostatic energy is a functional of the electron density $\rho^{\text {el}}(\boldsymbol{r})$. According to Eq. A2-4 in Ref. \citen{before_sccs_jcp_2009}, with the electrostatic energy given in the form of Eq.~\eqref{eq4}, the generalized Poisson equation with a dielectric tensor given in Eq.~\eqref{eq1}, and an assumption that the off-diagonal elements of the dielectric tensor are all zero, one can derive
\begin{equation}
\begin{aligned}
& \lim _{\lambda \rightarrow 0} \frac{E_{\left[\rho^{\text {el}}(\boldsymbol{r})+\lambda f(\boldsymbol{r})\right]}^\text{H}-E_{\left[\rho^{\text {el}}(\boldsymbol{r})\right]}^\text{H}}{\lambda} \\
& =\int\left(\phi^{\text{tot}}(\boldsymbol{r})-\frac{1}{8 \pi} \nabla \phi^{\text{tot}}(\boldsymbol{r}) \cdot  \left(\frac{\partial \boldsymbol{\epsilon}(\boldsymbol{r})}{\partial \rho^{\text {el}}(\boldsymbol{r})} \nabla \phi^{\text{tot}}(\boldsymbol{r})\right)\right) f(\boldsymbol{r}) d \boldsymbol{r} \\
& =\int\left(\phi^{\text{tot}}(\boldsymbol{r})-\frac{1}{8 \pi} \sum_{\text {i} \in\{\text {x,y,z}\}}\left(\nabla \phi^{\text{tot}}(\boldsymbol{r})\right)_\text{i}^2 \frac{\partial \epsilon_{\text {ii}}(\boldsymbol{r})}{\partial \rho^{\text {el}}(\boldsymbol{r})}\right) f(\boldsymbol{r}) d \boldsymbol{r}.
\label{eq13}
\end{aligned}
\end{equation}
As a result, one can find that the $E^\text {H}$’s contribution to the Kohn-Sham potential is
\begin{equation}
V^\text{H}(\boldsymbol{r})=\frac{\delta E^\text {H}}{\delta \rho^{\text {el}}(\boldsymbol{r})}=\phi^{\text{tot}}(\boldsymbol{r})-\frac{1}{8 \pi} \sum_{\text {i} \in\{\text {x,y,z}\}}\left(\nabla \phi^{\text{tot}}(\boldsymbol{r})\right)_\text{i}^2  \frac{\partial \epsilon_{\text {ii}}(\boldsymbol{r})}{\partial \rho^{\text {el}}(\boldsymbol{r})}.
\label{eq14}
\end{equation}
With the replacement of the dielectric function from a scalar to a tensor, the aforementioned contribution now involves the computation of the weighted Euclidean norm of $\nabla \phi^{\text{tot}}(\boldsymbol{r})$ rather than the standard Euclidean norm of $\nabla \phi^{\text{tot}}(\boldsymbol{r})$, as in Refs. \citen{before_sccs_jcp_2009, sccs_jcp_2012}. Similarly, the form of the analytical force contributed by $E^\text {H}$ can be derived as\cite{before_sccs_jcp_2009, sccs_jcp_2012, my_cpc_2025}
\begin{equation}
\begin{aligned}
& f_\text {i}^\text {A}=-\frac{\partial E^\text {H}}{\partial R_\text {i}^\text {A}}\\
&=-\frac{\partial \frac{1}{8 \pi} \int \nabla \phi^{\text{tot}}(\boldsymbol{r}) \cdot\left(\boldsymbol{\epsilon}(\boldsymbol{r}) \nabla \phi^{\text{tot}}(\boldsymbol{r})\right) d \boldsymbol{r}}{\partial R_\text {i}^\text {A}} \\
&=-\frac{1}{8 \pi}\left(\frac{\int\left(\nabla \partial \phi^{\text{tot}}(\boldsymbol{r})\right) \cdot \left(\boldsymbol{\epsilon}(\boldsymbol{r})  \nabla \phi^{\text{tot}}(\boldsymbol{r}) \right)d \boldsymbol{r}}{\partial R_\text {i}^\text {A}}+\frac{\int \nabla \phi^{\text{tot}}(\boldsymbol{r}) \cdot \left(\partial\boldsymbol{\epsilon}(\boldsymbol{r})  \nabla \phi^{\text{tot}}(\boldsymbol{r})\right) d \boldsymbol{r}}{\partial R_\text {i}^\text {A}}\right. \\
&\left.+\frac{\int \nabla \phi^{\text{tot}}(\boldsymbol{r}) \cdot \left(\boldsymbol{\epsilon}(\boldsymbol{r}) \nabla \partial \phi^{\text{tot}}(\boldsymbol{r}) \right)d \boldsymbol{r}}{\partial R_\text {i}^\text {A}}\right) \\
&=-\frac{1}{8 \pi}\left(\frac{2 \int \nabla \partial \phi^{\text{tot}}(\boldsymbol{r}) \cdot \left(\boldsymbol{\epsilon}(\boldsymbol{r})  \nabla \phi^{\text{tot}}(\boldsymbol{r})\right) d \boldsymbol{r}}{\partial R_\text {i}^\text {A}}+\frac{\int \nabla \phi^{\text{tot}}(\boldsymbol{r}) \cdot \left(\partial \boldsymbol{\epsilon}(\boldsymbol{r})  \nabla \phi^{\text{tot}}(\boldsymbol{r})\right) d \boldsymbol{r}}{\partial R_\text {i}^\text {A}}\right)  \\
&=-\frac{1}{8 \pi}\left(\frac{2 \int-\partial \phi^{\text{tot}}(\boldsymbol{r}) \nabla\cdot\left(\boldsymbol{\epsilon}(\boldsymbol{r}) \nabla \phi^{\text{tot}}(\boldsymbol{r})\right) d \boldsymbol{r}}{\partial R_\text {i}^\text {A}}+\frac{\int \nabla \phi^{\text{tot}}(\boldsymbol{r}) \cdot \left(\partial \boldsymbol{\epsilon}(\boldsymbol{r})  \nabla \phi^{\text{tot}}(\boldsymbol{r}) \right) d \boldsymbol{r}}{\partial R_\text {i}^\text {A}}\right)  \\
&=-\frac{1}{8 \pi}\left(\frac{8 \pi \int \partial \phi^{\text{tot}}(\boldsymbol{r})  \rho^{\text {solute}}(\boldsymbol{r}) d \boldsymbol{r}}{\partial R_\text {i}^\text {A}}+\frac{\int \nabla \phi^{\text{tot}}(\boldsymbol{r}) \cdot \left( \partial \boldsymbol{\epsilon}(\boldsymbol{r})  \nabla \phi^{\text{tot}}(\boldsymbol{r}) \right) d \boldsymbol{r}}{\partial R_\text {i}^\text {A}}\right)  \\
&=-\int \frac{\partial \phi^{\text{tot}}(\boldsymbol{r})}{\partial R_\text {i}^\text {A}}  \rho^{\text {solute}}(\boldsymbol{r}) d \boldsymbol{r}-\frac{1}{8 \pi} \int \nabla \phi^{\text{tot}}(\boldsymbol{r}) \cdot \left( \frac{\partial \boldsymbol{\epsilon}(\boldsymbol{r})}{\partial R_\text {i}^\text {A}} \nabla \phi^{\text{tot}}(\boldsymbol{r}) \right) d \boldsymbol{r}.
\label{eq15}
\end{aligned}
\end{equation}
$f_\text {i}^\text {A}$ and $R_\text {i}^\text {A}$ are denoted as the analytical force originated from $E^\text {H}$ acting on atom A in the i-direction (i$\in${x,y,z}) and the component of the position vector of atom A in the i-direction, respectively. From Eq.~\eqref{eq4}, we can derive
\begin{equation}
\begin{aligned}
f_\text {i}^\text {A} & =-\frac{\partial \frac{1}{2} \int \phi^{\text {tot}}(\boldsymbol{r}) \rho^{\text {solute}}(\boldsymbol{r}) d \boldsymbol{r}}{\partial R_\text {i}^\text {A}} \\
& =-\frac{1}{2} \int \frac{\partial \phi^{\text {tot}}(\boldsymbol{r})}{\partial R_\text {i}^\text {A}} \rho^{\text {solute}}(\boldsymbol{r}) d \boldsymbol{r}-\frac{1}{2} \int \phi^{\text {tot}}(\boldsymbol{r}) \frac{\partial \rho^{\text {solute}}(\boldsymbol{r})}{\partial R_\text {i}^\text {A}} d \boldsymbol{r}.
\label{eq16}
\end{aligned}
\end{equation}
Since the analytical forces given in Eqs.~\eqref{eq15} and~\eqref{eq16} are equal, one can establish the following equality:
\begin{equation}
\int \frac{\partial \phi^{\text {tot}}(\boldsymbol{r})}{\partial R_\text {i}^\text {A}}  \rho^{\text {solute}}(\boldsymbol{r}) d \boldsymbol{r}=\int \phi^{\text {tot}}(\boldsymbol{r}) \frac{\partial \rho^{\text {solute}}(\boldsymbol{r})}{\partial R_\text {i}^\text {A}} d \boldsymbol{r}-\frac{1}{4 \pi} \int \nabla \phi^{\text {tot}}(\boldsymbol{r}) \cdot \left(\frac{\partial \boldsymbol{\epsilon}(\boldsymbol{r})}{\partial R_\text {i}^\text {A}} \nabla \phi^{\text {tot}}(\boldsymbol{r})\right) d \boldsymbol{r}.
\label{eq17}
\end{equation}
By putting Eq.~\eqref{eq17} back to Eq.~\eqref{eq15}, one has
\begin{equation}
\begin{gathered}
f_\text {i}^\text {A}=\frac{1}{8 \pi} \int \nabla \phi^{\text {tot}}(\boldsymbol{r}) \cdot \left(\frac{\partial \boldsymbol{\epsilon}(\boldsymbol{r})}{\partial R_\text {i}^\text {A}}  \nabla \phi^{\text {tot}}(\boldsymbol{r}) \right) d \boldsymbol{r}-\int \phi^{\text {tot}}(\boldsymbol{r}) \frac{\partial \rho^{\text {solute}}(\boldsymbol{r})}{\partial R_\text {i}^\text {A}} d \boldsymbol{r} \\
=\frac{1}{8 \pi} \int \sum_{\text {j} \in\{\text {x,y,z}\}} \frac{\partial \epsilon_{\text {jj}}(\boldsymbol{r})}{\partial R_\text {i}^\text {A}}\left(\nabla \phi^{\text {tot}}(\boldsymbol{r})\right)_\text {j}^2 d \boldsymbol{r}-\int \phi^{\text {tot}}(\boldsymbol{r}) \frac{\partial \rho^{\text {solute}}(\boldsymbol{r})}{\partial R_\text {i}^\text {A}} d \boldsymbol{r} \\
=\frac{1}{8 \pi} \int \sum_{\text {j} \in\{\text {x,y,z}\}} \frac{\partial \epsilon_{\text {jj}}(\boldsymbol{r})}{\partial R_\text {i}^\text {A}}\left(\nabla \phi^{\text {tot}}(\boldsymbol{r})\right)_\text {j}^2 d \boldsymbol{r}-\int \phi^{\text {tot}}(\boldsymbol{r}) \frac{\partial \rho^{\text {el}}(\boldsymbol{r})}{\partial R_\text {i}^\text {A}} d \boldsymbol{r}-\int \phi^{\text {tot}}(\boldsymbol{r}) \frac{\partial n_\text {c}^\text {A}(\boldsymbol{r})}{\partial R_\text {i}^\text {A}} d \boldsymbol{r} .
\label{eq18}
\end{gathered}
\end{equation}
$n_\text {c}^\text {A}(\boldsymbol{r})$ is the effective charge density of ion A which represents the total charge distribution of the nucleus of A and the electrons of atom A that are not explicitly treated by basis functions\cite{cp2k_quickstep_cpc_2005}. In the region where $\rho^{\text {el}}(\boldsymbol{r})>\rho_{\min}$, the dependence of $\epsilon_{\text {ii}}(\boldsymbol{r})$ on $R_\text {i}^\text {A}$ via $\rho^{\text {el}}(\boldsymbol{r})$ is the same as that in the standard SCCS method. With reference to Eq. 10 in Ref. \citen{my_cpc_2025}, we have
\begin{equation}
\begin{aligned}
& \frac{1}{8 \pi} \int_{\rho^{\text {el}}(\boldsymbol{r})>\rho_{\min}} \sum_{\text {j} \in\{\text {x,y,z}\}} \frac{\partial \epsilon_{\text {jj}}(\boldsymbol{r})}{\partial R_\text {i}^\text {A}}\left(\nabla \phi^{\text{tot}}(\boldsymbol{r})\right)_\text {j}^2 d \boldsymbol{r} \\
& =\frac{1}{8 \pi} \int_{\rho^{\text {el}}(\boldsymbol{r})>\rho_{\min}} \sum_{\text {j} \in\{\text {x,y,z}\}} \frac{\partial \epsilon_{\text {jj}}(\boldsymbol{r})}{\partial \rho^{\text {el}}(\boldsymbol{r})} \frac{\partial \rho^{\text {el}}(\boldsymbol{r})}{\partial R_\text {i}^\text {A}}\left(\nabla \phi^{\text{tot}}(\boldsymbol{r})\right)_\text {j}^2 d \boldsymbol{r} \\
& =\frac{1}{8 \pi} \int \sum_{\text {j} \in\{\text {x,y,z}\}}\left(\nabla \phi^{\text{tot}}(\boldsymbol{r})\right)_\text {j}^2 \frac{\partial \epsilon_{\text {jj}}(\boldsymbol{r})}{\partial \rho^{\text {el}}(\boldsymbol{r})} \frac{\partial \rho^{\text {el}}(\boldsymbol{r})}{\partial R_\text {i}^\text {A}} d \boldsymbol{r} .
\label{eq19}
\end{aligned}
\end{equation}
In the region where $\rho^{\text {el}}(\boldsymbol{r})<\rho_{\min}$, $\epsilon_{\text {jj}}(\boldsymbol{r})$ equals $\epsilon_{\text {dist,jj}}(\boldsymbol{r})$ and obeys the z-direction spatial distribution directly depending only on $r_\text {z}^{\text {ref}}$ but not relying on $\rho^{\text {el}}(\boldsymbol{r})$. We here consider that $r_\text {z}^{\text {ref}}$ is the average of the z-components of coordinates of the outermost atomic layer of a solid surface. Based on the formulas given in Eq.~\eqref{eq9}, we can derive $\frac{\partial \epsilon_{\text {jj}}(\boldsymbol{r})}{\partial R_\text {i}^\text {A}}$ in Eq.~\eqref{eq18} as
\begin{equation}
\begin{aligned}
\frac{\partial \epsilon_{\text {jj}}(\boldsymbol{r})}{\partial R_\text {i}^\text {A}} & =\frac{\partial \epsilon_{\text {dist,jj}}(\boldsymbol{r})}{\partial R_\text {i}^\text {A}} \\
& =\frac{\partial\left(\epsilon_{\text {dist,jj}}^{\text {near}}+0.5  \left(\epsilon_{\text {bulk}}-\epsilon_{\text {dist,jj}}^{\text {near}}\right)  \left(1+\operatorname{erf}\left(\frac{d-d_\text {j}^\text {0}}{\sigma}\right)\right)\right)}{\partial R_\text {i}^\text {A}} \\
& =\left(\epsilon_{\text {bulk}}-\epsilon_{\text {dist,jj}}^{\text {near}}\right)   \frac{1}{\sqrt{\pi}}   e^{-\left(\frac{d-d_\text {j}^\text {0}}{\sigma}\right)^2} \frac{\partial d}{\partial R_\text {i}^\text {A}} .
\label{eq20}
\end{aligned}
\end{equation}
In the case of Eq.~\eqref{eq10}, we define $X=r_\text {z}-r_\text {z}^\text {ref}-d_\text {j}^\text {0}$ for simplifying the notations, and one can derive $\frac{\partial d}{\partial R_\text {i}^\text {A}}$ as
\begin{equation}
\begin{aligned}
\frac{\partial d}{\partial R_\text {i}^\text {A}} & =\left(1+\frac{1+\alpha \tanh (\beta X)-\alpha \beta X(\operatorname{sech}(\beta X))^2}{(1+\alpha \tanh (\beta X))^2}\right) \frac{\partial X}{\partial R_\text {i}^\text {A}} \\
& =\left(1+\frac{1+\alpha \tanh (\beta X)-\alpha \beta X(\operatorname{sech}(\beta X))^2}{(1+\alpha \tanh (\beta X))^2}\right)\left(-\frac{1}{N_{\text {atoms}}^{\text {ref}}}\right) .
\label{eq21}
\end{aligned}
\end{equation}
$N_{\text {atoms}}^{\text {ref}}$ is the number of atoms in the outermost atomic layer of the solid surface model. In the case of Eq.~\eqref{eq11}, we have $X=r_\text {z}^\text {ref}-r_\text {z}-d_\text {j}^\text {0}$ and
\begin{equation}
\frac{\partial d}{\partial R_\text {i}^\text {A}}=\left(1+\frac{1+\alpha \tanh (\beta X)-\alpha \beta X(\operatorname{sech}(\beta X))^2}{(1+\alpha \tanh (\beta X))^2}\right)\left(\frac{1}{N_{\text {atoms}}^{\text {ref}}}\right) .
\label{eq22}
\end{equation}
$\frac{\partial \epsilon_{\text {jj}}(\boldsymbol{r})}{\partial R_\text {i}^\text {A}}$ now can be calculated by following Eqs.~\eqref{eq20}--\eqref{eq21} or Eqs.~\eqref{eq20} and~\eqref{eq22}. The analytical forces given by Eq.~\eqref{eq18} can then be calculated analytically, contributed from both of the regions of dielectric functions $\epsilon_{\text {jj}}(\boldsymbol{r}) (\text {j} \in\{\text {x,y,z}\})$ where $\rho^{\text {el}}(\boldsymbol{r})<\rho_{\min}$ and $\rho^{\text {el}}(\boldsymbol{r})>\rho_{\min}$,
\begin{equation}
\begin{aligned}
f_\text {i}^\text {A}= & \frac{1}{8 \pi} \int \sum_{\text {j} \in\{\text {x,y,z}\}}\left(\frac{\partial \epsilon_{\text {jj}}(\boldsymbol{r})}{\partial \rho^{\text {el}}(\boldsymbol{r})} \frac{\partial \rho^{\text {el}}(\boldsymbol{r})}{\partial R_\text {i}^\text {A}}+\left(\epsilon_{\text{bulk}}-\epsilon_{\text {dist,jj}}^{\text{near}}\right)   \frac{1}{\sqrt{\pi}}   e^{-\left(\frac{d-d_\text {j}^\text {0}}{\sigma}\right)^2} \frac{\partial d}{\partial R_\text {i}^\text {A}}\right)\left(\nabla \phi^{\text{tot}}(\boldsymbol{r})\right)_\text {j}^2 d \boldsymbol{r} \\
& -\int \phi^{\text{tot}}(\boldsymbol{r}) \frac{\partial \rho^{\text {el}}(\boldsymbol{r})}{\partial R_\text {i}^\text {A}} d \boldsymbol{r}-\int \phi^{\text{tot}}(\boldsymbol{r}) \frac{\partial n_\text {c}^\text {A}(\boldsymbol{r})}{\partial R_\text {i}^\text {A}} d \boldsymbol{r} \\
= & -\int\left(\phi^{\text{tot}}(\boldsymbol{r})-\frac{1}{8 \pi} \sum_{\text {j} \in\{\text {x,y,z}\}}\left(\nabla \phi^{\text{tot}}(\boldsymbol{r})\right)_\text {j}^2 \frac{\partial \epsilon_{\text {jj}}(\boldsymbol{r})}{\partial \rho^{\text {el}}(\boldsymbol{r})}\right) \frac{\partial \rho^{\text {el}}(\boldsymbol{r})}{\partial R_\text {i}^\text {A}} d \boldsymbol{r}-\int \phi^{\text{tot}}(\boldsymbol{r}) \frac{\partial n_\text {c}^\text {A}(\boldsymbol{r})}{\partial R_\text {i}^\text {A}} d \boldsymbol{r} \\
& +\frac{1}{8 \pi} \int \sum_{\text {j} \in\{\text {x,y,z}\}}\left(\epsilon_{\text{bulk}}-\epsilon_{\text {dist,jj}}^{\text{near}}\right)   \frac{1}{\sqrt{\pi}}   e^{-\left(\frac{d-d_\text {j}^\text {0}}{\sigma}\right)^2} \frac{\partial d}{\partial R_\text {i}^\text {A}}\left(\nabla \phi^{\text{tot}}(\boldsymbol{r})\right)_\text {j}^2 d \boldsymbol{r} \\
& =-\int V^\text{H}(\boldsymbol{r}) \frac{\partial \rho^{\text {el}}(\boldsymbol{r})}{\partial R_\text {i}^\text {A}} d \boldsymbol{r}-\int \phi^{\text{tot}}(\boldsymbol{r}) \frac{\partial n_\text {c}^\text {A}(\boldsymbol{r})}{\partial R_\text {i}^\text {A}} d \boldsymbol{r} \\
& +\frac{1}{8 \pi} \int \sum_{\text {j} \in\{\text {x,y,z}\}}\left(\epsilon_{\text{bulk}}-\epsilon_{\text {dist,jj}}^{\text{near}}\right)   \frac{1}{\sqrt{\pi}}   e^{-\left(\frac{d-d_\text {j}^\text {0}}{\sigma}\right)^2} \frac{\partial d}{\partial R_\text {i}^\text {A}}\left(\nabla \phi^{\text{tot}}(\boldsymbol{r})\right)_\text {j}^2 d \boldsymbol{r} .
\label{eq23}
\end{aligned}
\end{equation}
\section{3. Finite-Eelement Solver for the Anisotropic Poisson Equation}
\begin{figure}[H]
\centering
\includegraphics[width=\textwidth]{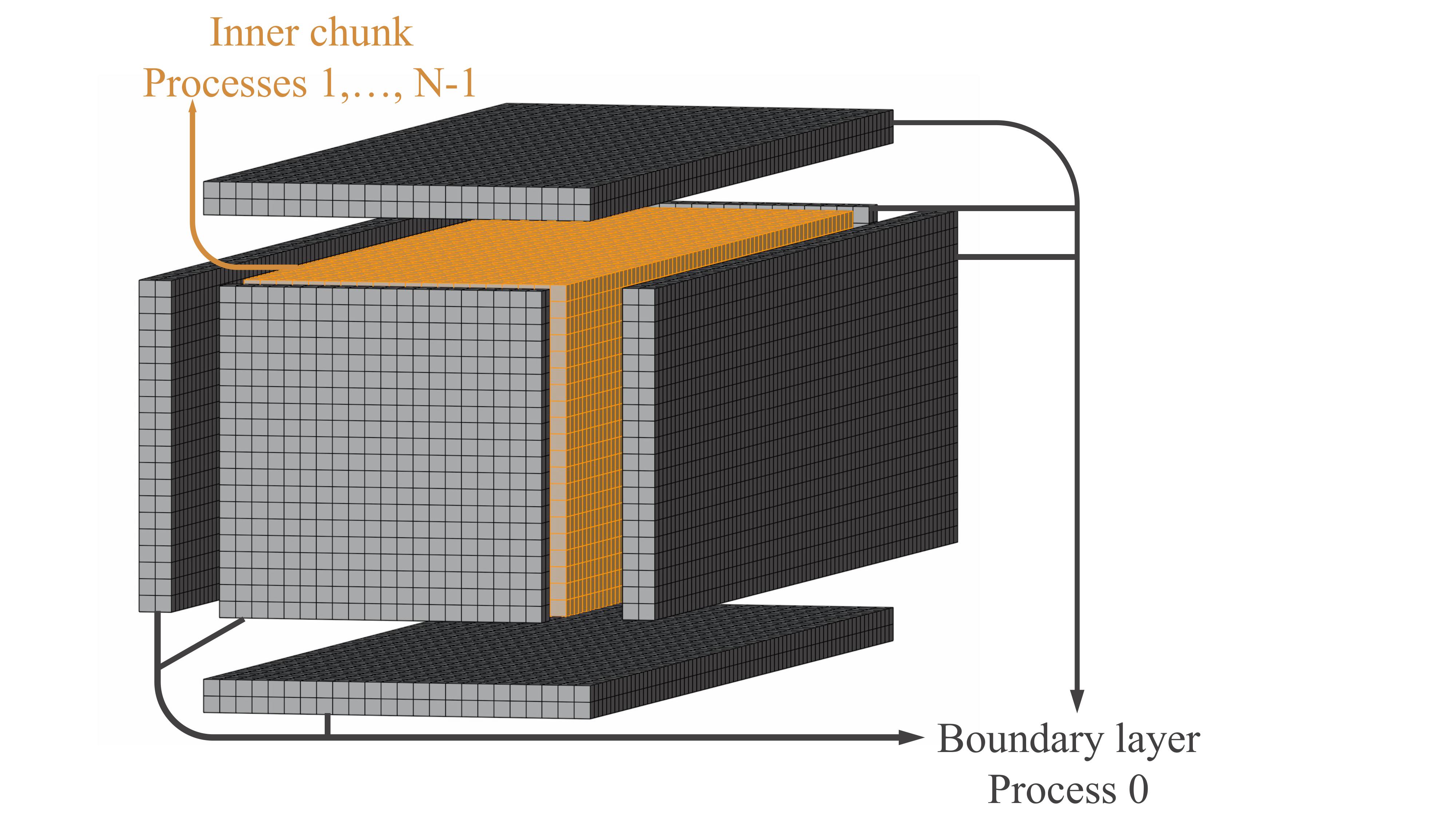}
\caption{The schematic of the hexahedron mesh and parallel assignment strategy in MPI. The outermost gray hexahedral elements are assigned to the same process, “0”. The remaining elements are evenly distributed among the remaining N-1 processes, “1, …, N-1”. The total number of processes is N.}\label{fig1}
\end{figure}

In the anisotropic generalized Poisson equation as formulated in Eq.~\eqref{eq1}, the three diagonal components of the dielectric tensor are multiplicatively coupled with $\frac{\partial^2 \phi^{\text{tot}}(\boldsymbol{r})}{\partial r_\text {i}^2}($for $\text {i} \in\{\text {x,y,z}\})$. As a result, the reformulation of the isotropic Poisson equation in SCCS (as given by Eqs. 8 and 9 in Ref. \citen{ sccs_jcp_2012}) and its solution via Fast Fourier Transform (FFT) and an iterative method are no longer applicable for solving Eq.~\eqref{eq1}. The finite element method provides a viable approach for solving anisotropic generalized Poisson equations.\\
FEniCSx\cite{BarattaEtal2023, ScroggsEtal2022, BasixJoss, AlnaesEtal2014} is a popular open-source computing platform for solving partial differential equations with the finite element method. DOLFINx\cite{BarattaEtal2023} is the computational environment of FEniCSx and implements the FEniCSx Problem Solving Environment in C++ and Python. A solver for the anisotropic generalized Poisson equation (Eq.~\eqref{eq1}) was developed using Python and DOLFINx, and an interface was implemented to couple it with the CP2K software package. This interface enables direct exchange of memory pointers between CP2K and the solver, including those for the charge density and dielectric tensor, and allows CP2K to make use of the solution $
\phi^{\text{tot}}(\boldsymbol{r})$ to Eq.~\eqref{eq1}, the core equation of the AICS model.\\
A three-dimensional uniform finite element mesh is constructed using linear Lagrange basis functions. A representative schematic of the hexahedron mesh, along with its distribution across processes, is presented in Fig.~\ref{fig1}. The constructed mesh matches the shape, size, and subdivision of the real-space grid in CP2K. The two outermost layers (the grey elements in Fig.~\ref{fig1}) of hexahedral elements, are assigned to a single process (process 0). The rest of the hexahedral elements (the red elements in Fig.~\ref{fig1}) are evenly distributed among the remaining processes. This partition is to facilitate the construction of three-dimensional periodic boundary conditions for the mesh while ensuring good computational efficiency.\\
The vertices on the outermost surface of the hexahedral mesh can be categorized into those within the 6 faces of the box (excluding boundary points), those along the 12 edges of the box (excluding boundary points), and those at the 8 vertices of the box. Due to the symmetrical  constraints of the three-dimensional periodic boundary conditions, only the vertices on three faces of the box (excluding boundary points), three edges (excluding boundary points), and one vertex need to be retained. The remaining vertices can be generated based on these reference points and the box dimensions. To impose three-dimensional periodic boundary conditions for linear Lagrange hexahedral elements, equivalent vertices on opposite boundary facets are identified and collapsed to a single global vertex index. The cell-to-vertex adjacency is then rebuilt using the updated vertex indices. Since all degrees of freedom in the CG1 element reside at vertices, this vertex-level identification automatically induces periodic equivalence of edges, faces, and cells, without the need for explicit edge-to-edge or face-to-face mappings. The above procedure to create a periodic mesh followed an example provided on the FEniCS user forum by Dokken\cite{fenics_forum_pbc}.\\ 
At each SCF step, the dielectric functions $\epsilon_{\text {ii}}(\boldsymbol{r})(\text {i} \in\{\text {x,y,z}\})$ are computed from the updated electron density $\rho^{\text {el}}(\boldsymbol{r})$ according to Eqs.~\eqref{eq5}--\eqref{eq11}, and the solute’s charge density $\rho^{\text {solute}}(\boldsymbol{r})$ on the same real-space grid is computed as the sum of $\rho^{\text {el}}(\boldsymbol{r})$ and $n_\text {c}^\text {A}(\boldsymbol{r})$. These real-space grid data in CP2K are then mapped onto the three-dimensional periodic finite element mesh. In practice, the values from CP2K’s grid are directly assigned to the degrees of freedom of the finite element space based on coordinate matching. Once mapped, the finite element basis functions naturally represent the charge density and dielectric functions, allowing us to proceed with the subsequent finite element calculations in a consistent manner.\\
In DOLFINx, one typically starts with a partial differential equation in its variational (weak) form. Keeping only the principal diagonal components of the dielectric tensor in Eq.~\eqref{eq2}, Eq.~\eqref{eq1} can be rewritten in the following weak form for solution\\
\begin{equation}
\int \frac{\partial \phi^{\text{tot}}(\boldsymbol{r})}{\partial r_\text {x}} \frac{\partial v(\boldsymbol{r})}{\partial r_\text {x}} \epsilon_{\text {xx}}(\boldsymbol{r})+\frac{\partial \phi^{\text{tot}}(\boldsymbol{r})}{\partial r_\text {y}} \frac{\partial v(\boldsymbol{r})}{\partial r_\text {y}} \epsilon_{\text {yy}}(\boldsymbol{r})+\frac{\partial \phi^{\text{tot}}(\boldsymbol{r})}{\partial r_\text {z}} \frac{\partial v(\boldsymbol{r})}{\partial r_\text {z}} \epsilon_{\text {zz}}(\boldsymbol{r}) d \boldsymbol{r}=4 \pi \int v(\boldsymbol{r}) \rho^{\text {solute}}(\boldsymbol{r}) d \boldsymbol{r}.
\label{eq24}
\end{equation}
The weak form can be assembled into a linear algebraic system. The iterative solver was configured using PETSc, with the Conjugate Gradient (CG) method as the default Krylov subspace solver (ksp\_type = cg). The relative and absolute tolerances were set to $10^{-10}$ (ksp\_rtol) and $10^{-12}$ (ksp\_atol), respectively, with a maximum iteration limit of 100,000 (ksp\_max\_it). The Geometric-Algebraic Multigrid (GAMG) preconditioner (pc\_type = gamg) was applied to improve convergence. After the electrostatic potential is obtained in the form of a finite element function, the electrostatic potential values at the mesh vertices are accordingly transferred to CP2K's real-space grid for subsequent computations within CP2K.\\
We have developed a C/Fortran interface that allows CP2K to call a Python script that makes use of DOLFINx. A C bridge is used to connect the C function call in a Fortran subroutine of CP2K to the execution call of the Python/DOLFINx Poisson solver program in the C bridge. To ensure compatibility between FEniCSx/DOLFINx and CP2K, both packages were built on the HPC system using the same set of dependency libraries, which were chosen to be consistent with the cluster’s MPI environment, thus enabling correct MPI-based parallel execution of CP2K and FEAPS.\\
\section{4. Results and Discussion}
\subsection{4.1 Testing of the Analytical Forces Contributed by $E^\text {H}$}
\begin{figure}[H]
\centering
\includegraphics[width=\textwidth]{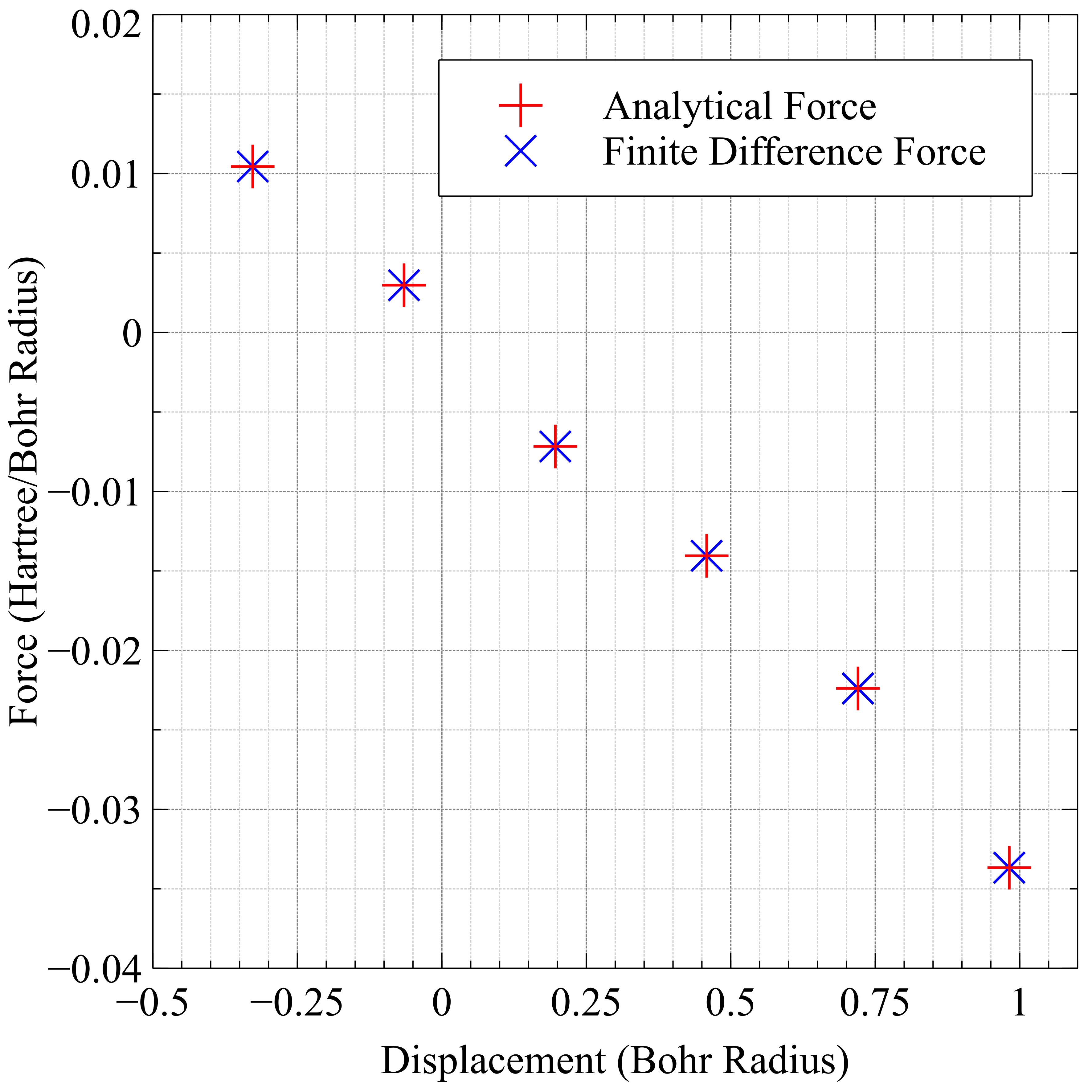}
\caption{Calculated analytical (red “+” symbols) and finite-difference (blue “×” symbols) forces on the Ag atom in the outermost atomic layer across a series of displacements.}\label{fig2}
\end{figure}

It is essential to ensure the accuracy of the computed analytical forces. We validate the consistency between the total energy and the analytical force which follows the formulas given in Sec. 2.2.3, and calculated by using DFT + the FEAPS-based AICS model that we developed and implemented in the CP2K software package. A cubic 108‑atom bulk Ag supercell was first optimized using the PBE\cite{pbe} functional to obtain relaxed lattice parameters and atomic coordinates. A five-atomic-layer (4$\times$4) Ag(111) slab was then cleaved from this optimized bulk structure and placed at the center of the rectangular box. The simulation box is periodic in all three dimensions, with a solvent buffer of approximately 24 \AA\ introduced between adjacent Ag slabs. The slab model in the .cif format is provided in Sec. SI.F and the side view is presented in Fig.~\ref{figs1}. The slab was embedded in the anisotropic implicit water, where the parameters given in detail in Sec. 4.2.3 were used. $\rho_{\max}$ and $\rho_{\min}$ in Eq.~\eqref{eq5} were set to 0.001 and 0.0001 (unit:  ${\text{electrons/(Bohr Radius)}}^3$; this unit will be used for all electron density values mentioned in the rest of the paper and we will not explicitly mention the unit anymore), respectively, which ensures high accuracy in the finite-difference force evaluations, as evidenced by the sufficiently small differences between the forward and backward energy slopes. To ensure that no unphysical implicit solvent region exists inside the Ag(111) slab (as reported in Refs.\citen{solvent_aware_jctc_2019,my_cpc_2025}), the values of the dielectric functions within the slab were manually set to 1. Six geometries were subsequently generated by displacing a Ag atom in the outermost layer of the slab along the (1, 1, 1) direction, with displacement magnitudes ranging from -0.33 Bohr to 0.98 Bohr. For each of these configurations, a DFT+AICS calculation of total energy and forces was performed. All calculations employed a Fermi–Dirac smearing corresponding to a temperature of 300 K and an SCF convergence threshold of $1.0 \times 10^{-9}$ (keyword: EPS\_SCF). The parameters for the finite element Poisson solver followed those reported in Sec. 3. The DZVP-MOLOPT-SR-GTH-PBE-q11 basis sets\cite{molopt_basis} and GTH-PBE-q11 pseudopotentials\cite{gth_pp1,gth_pp2,gth_pp3} were adopted, and the planewave cutoff at the finest level of the multigrid was set to 320 Ry. The PBE functional was used as the exchange correlation functional to perform all calculations.\\
For each of the displaced structures described above, two additional small displacements (in the magnitude of $1.0 \times 10^{-5}$ Bohr) were applied to the perturbed atom along the direction of the total atomic force acting on it, and DFT+AICS total energy calculations were then performed for the two structures. The forward and backward energy slopes with respect to the additional small displacement were first computed and then averaged to get the finite difference force. Fig.~\ref{fig2} presents the calculated analytical and finite-difference forces acting on the perturbed atom at a series of displacements of the atom, and their differences and the ratios between force difference and analytical force are given in Table~\ref{tab1}.\\
As shown in Fig.~\ref{fig2}, the analytical forces show good agreement with those obtained from finite-difference calculations over the range of displacements. The maximal absolute force difference is $7.1 \times 10^{-7}$ Hartree/Bohr as revealed in Table~\ref{tab1}. As expected, the aforementioned ratios are larger if the analytical forces are smaller. The largest absolute percentage is 0.0238\% which resulted from the lowest analytical force 0.003 Hartree/Bohr at -0.07 Bohr displacement.\\
\begin{table}[H]
    \centering
    \caption{Differences between analytical and finite-difference forces, and percentages of these differences relative to the analytical forces.}
    \label{tab1}

\begin{tabular}{| l | l | l |}
\hline
Displacement (Bohr Radius) & Difference (Hartree/Bohr Radius) & Percentage \\
\hline
-0.33 & 3.6×10\textsuperscript{-8}& 0.0003\%\\
\hline
-0.07 & -7.1×10\textsuperscript{-7}& -0.0238\%\\
\hline
0.20 & 6.7×10\textsuperscript{-7}& -0.009\%\\
\hline
0.46 & 4.9×10\textsuperscript{-7}& -0.004\%\\
\hline
0.72 & 3.7×10\textsuperscript{-7}& -0.002\%\\
\hline
0.98 & 3.3×10\textsuperscript{-8}& -0.0001\%\\
\hline

\end{tabular}
\end{table}
\subsection{4.2 Electrostatic Potential}
In the previous section, we verified the consistency between the mathematical formulations of electrostatic energy and analytic forces and the correctness of their implementation in the codes. In this section, we present the electrostatic potential calculated by using AICS method and FEAPS to further verify and test them.\\
\subsubsection{4.2.1 Vacuum}
First, we excluded the influence of implicit solvent and verified the correctness of FEAPS alone. We set all dielectric functions to 1 throughout the entire box and performed DFT+AICS\&FEAPS SCF energy calculations on the Ag(111) surface slab model (see Sec. SI.C for details). With this computational setup, we were effectively performing a DFT SCF calculation for the Ag(111) surface slab model in the vacuum environment. The isosurfaces of the electrostatic potential calculated at SCF convergence are shown in Fig.~S\ref{figs2a}. For comparison, the iso-surface of the electrostatic potential at SCF convergence calculated using the existing FFT-based Poisson solver in CP2K is also shown in Fig.~S\ref{figs2b}. The two iso-surfaces are visually almost identical, as shown in these figures. The variation of the planar averages of the above two electrostatic potentials, averaged within the plane parallel to the Ag(111) surface, along the surface normal direction is shown in Fig.~S\ref{figs2.5a}. As it is shown in the figure, the results given by FEAPS and the existing FFT-based Poisson solver are in good agreement.\\
The maximum absolute difference of the values of the above-mentioned electrostatic potentials at the same grid element is 0.06 Hartree. These grid elements were located near the atomic nuclei. The relatively large discrepancy originates from the difference in function representations: our finite element solver interpolates function value at a given position within each grid element linearly based on the values at the eight vertices of each grid element, while the FFT-based Poisson solver assumes a uniform distribution of the function within each element. The discrepancy is more pronounced near the nuclei, where the charge density of pseudopotentials exhibits rapid spatial variation. To examine the mean absolute difference between the two electrostatic potentials at all of the grid elements, we found an average value of $7.0 \times 10^{-4}$ Hartree, indicating a good average agreement between the two real-space potential functions.\\
The work function, defined and calculated by using Eq. 2 in Ref. \citen{our_gce_jctc_2024}, is the so‑called potential of zero charge because our slab was charge neutral. The finite-element Poisson solver gave the value 4.18 eV, and the traditional FFT-based solver resulted in the value 4.14 eV. This discrepancy (0.04 eV) in work functions might be due to the above-mentioned difference in electrostatic potentials in the vicinity of the atomic nuclei. Taking into account the fundamental algorithmic differences between first-order linear finite elements and the FFT-based solver, we consider this discrepancy to be reasonable.\\
\subsubsection{4.2.2 Isotropic}
Next, we examined the consistency between the electrostatic potential computed by using AICS\&FEAPS, in the computational setup where the dielectric is isotropic and spatially uniform in the region of $\rho_{\min}>\rho^{\text {el}}(\boldsymbol{r})$, and the one obtained from the conventional SCCS method. In the former case, we set $\epsilon_{\text {dist,xx}}^{\text {near}}, \epsilon_{\text {dist,yy}}^{\text {near}}, \epsilon_{\text {dist,zz}}^{\text {near}}$, and $\epsilon_{\text {bulk}}$ to 78.3 (relative dielectric constant of bulk water under room temperature) to achieve isotropy and spatial uniformity in the region of $\rho_{\min}>\rho^{\text {el}}(\boldsymbol{r})$. In the latter case, we set $\epsilon_{\text {bulk}}$ to 78.3. In both cases, we used $\rho_{\max}=0.01$ and $\rho_{\min}=0.001$ as solute-solvent boundary parameters. Figs.~S\ref{figs2c} and~S\ref{figs2d} present the iso-surfaces of the calculated electrostatic potentials at SCF convergence in the above two cases, respectively. It is shown that the two iso-surfaces appear nearly identical. Similarly, the plots of the planar-averaged electrostatic potentials, computed by averaging over planes parallel to the Ag(111) surface, as a function of the coordinate perpendicular to the surface are also in good agreement too, as shown in Fig.~S\ref{figs2.5b}.\\
The maximum absolute difference between the two electrostatic potentials at the same grid element is 0.06 Hartree, with the largest deviations observed near atomic nuclei, which is the same as what was observed in the vacuum calculations. The mean absolute difference across all grid elements is around $6 \times 10^{-4}$ Hartree, indicating overall good agreement between the two real-space electrostatic potentials.\\
The work functions calculated by our model powered by finite element–based Poisson solver and the SCCS model are 3.44 eV and 3.41 eV, respectively. The difference is around 0.03 eV, and is consistent with what was observed in the previous vacuum test, presumably caused by the discrepancy between the algorithm foundations of the finite element and FFT-based Poisson solvers. The well-documented work function reduction as the slab was embedded in an implicit solvent environment\cite{hormann_gce,our_gce_jctc_2024} was observed in the case of our model powered by FEAPS. It was also observed in the SCCS case as expected.\\
\subsubsection{4.2.3 Anisotropic}
\begin{figure}[H]
\centering
\includegraphics[width=\textwidth]{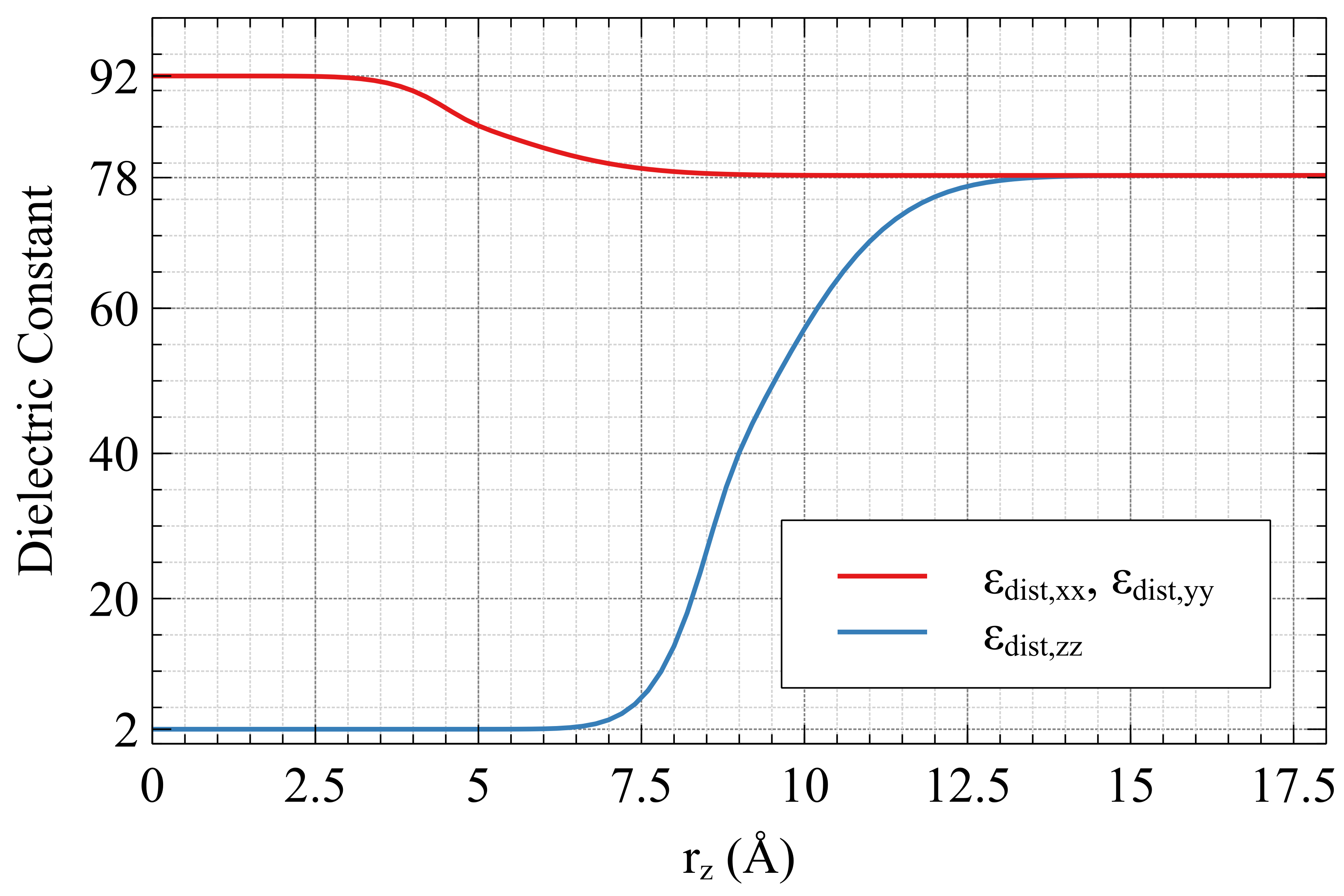}
\caption{Profiles of $\epsilon_{\text {dist,xx}}^{\text {near}}, \epsilon_{\text {dist,yy}}^{\text {near}}$ (red), and $\epsilon_{\text {dist,zz}}^{\text {near}}$ (blue) as a function of $r_\text {z}$. The dielectric functions were defined by Eqs.~\eqref{eq9}--\eqref{eq10}. The z-coordinate reference $r_\text {z}^\text {ref}$ was set to 0.}\label{fig3}
\end{figure}

\begin{figure}[H]
\centering
\includegraphics[width=\textwidth]{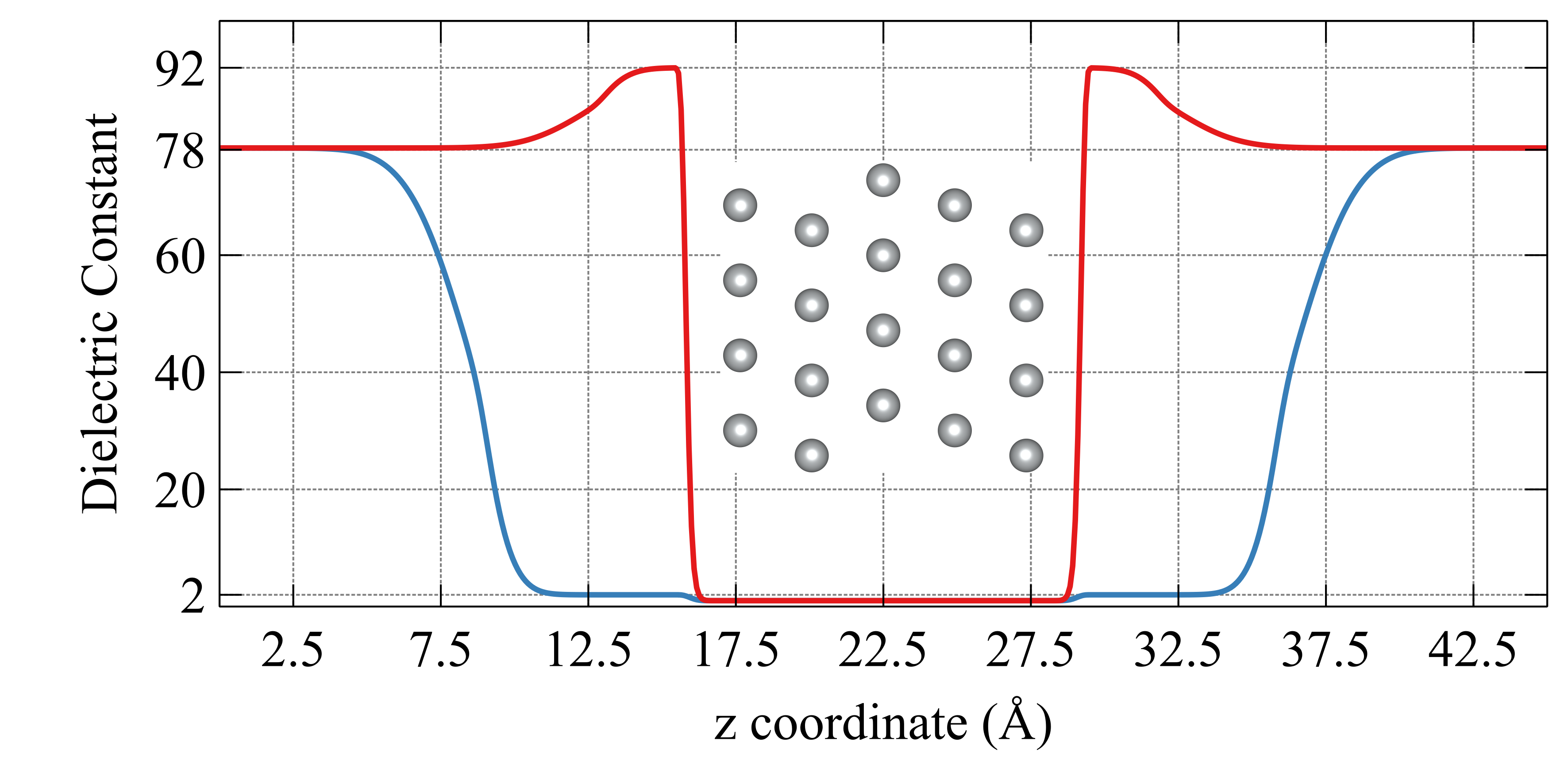}
\caption{Profiles of the planar-averaged values of in-plane ($\epsilon_{\text {dist,xx}}^{\text {near}}$ or $\epsilon_{\text {dist,yy}}^{\text {near}}$, in red) and out-of-plane ($\epsilon_{\text {dist,zz}}^{\text {near}}$, in blue) dielectric functions as functions of z coordinates.}\label{fig4}
\end{figure}

\begin{figure}[H]
\centering

\subfigure[]{
\includegraphics[width=0.7\textwidth]{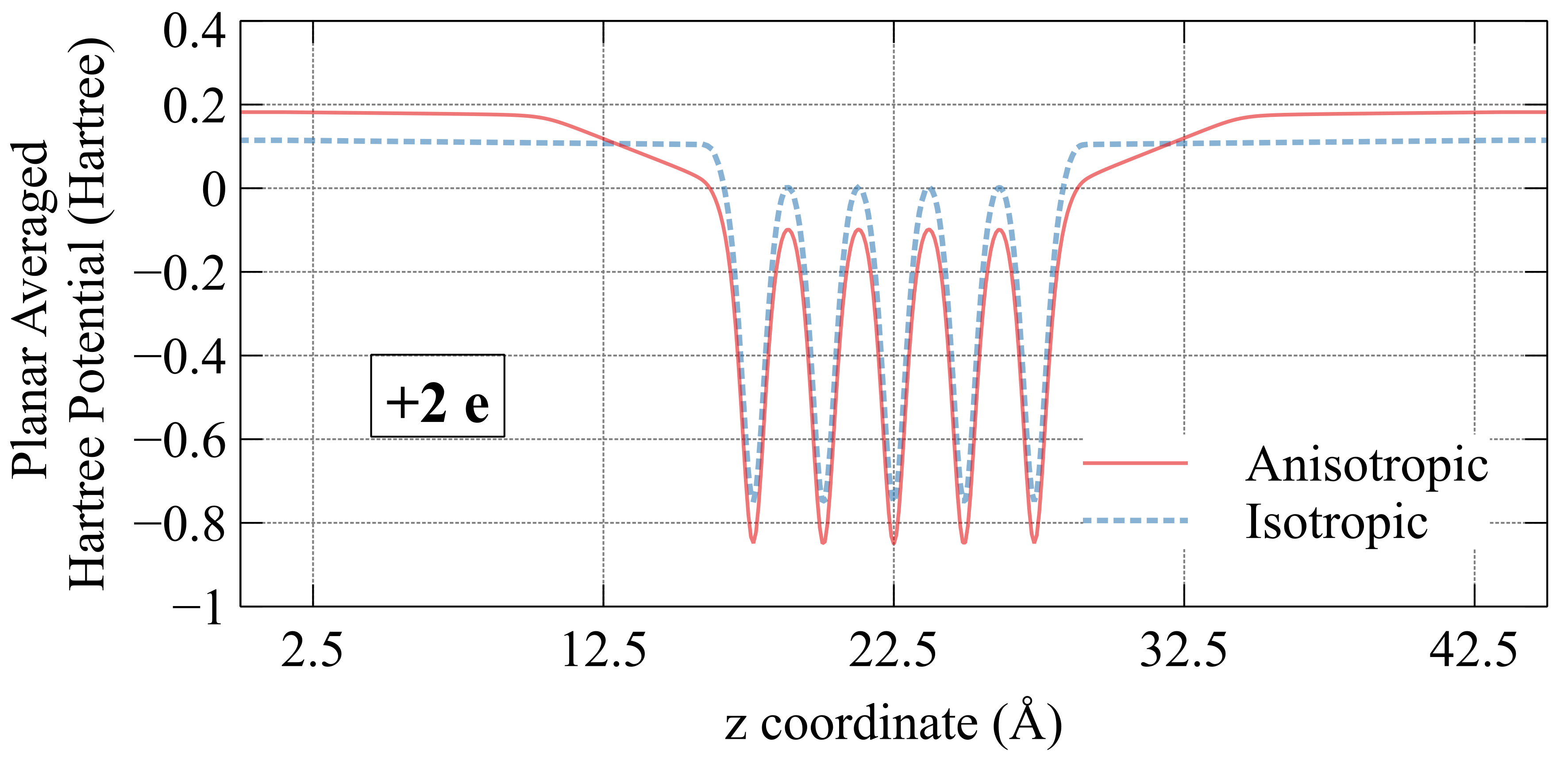}
\label{fig5a}
}
\subfigure[]{
\includegraphics[width=0.7\textwidth]{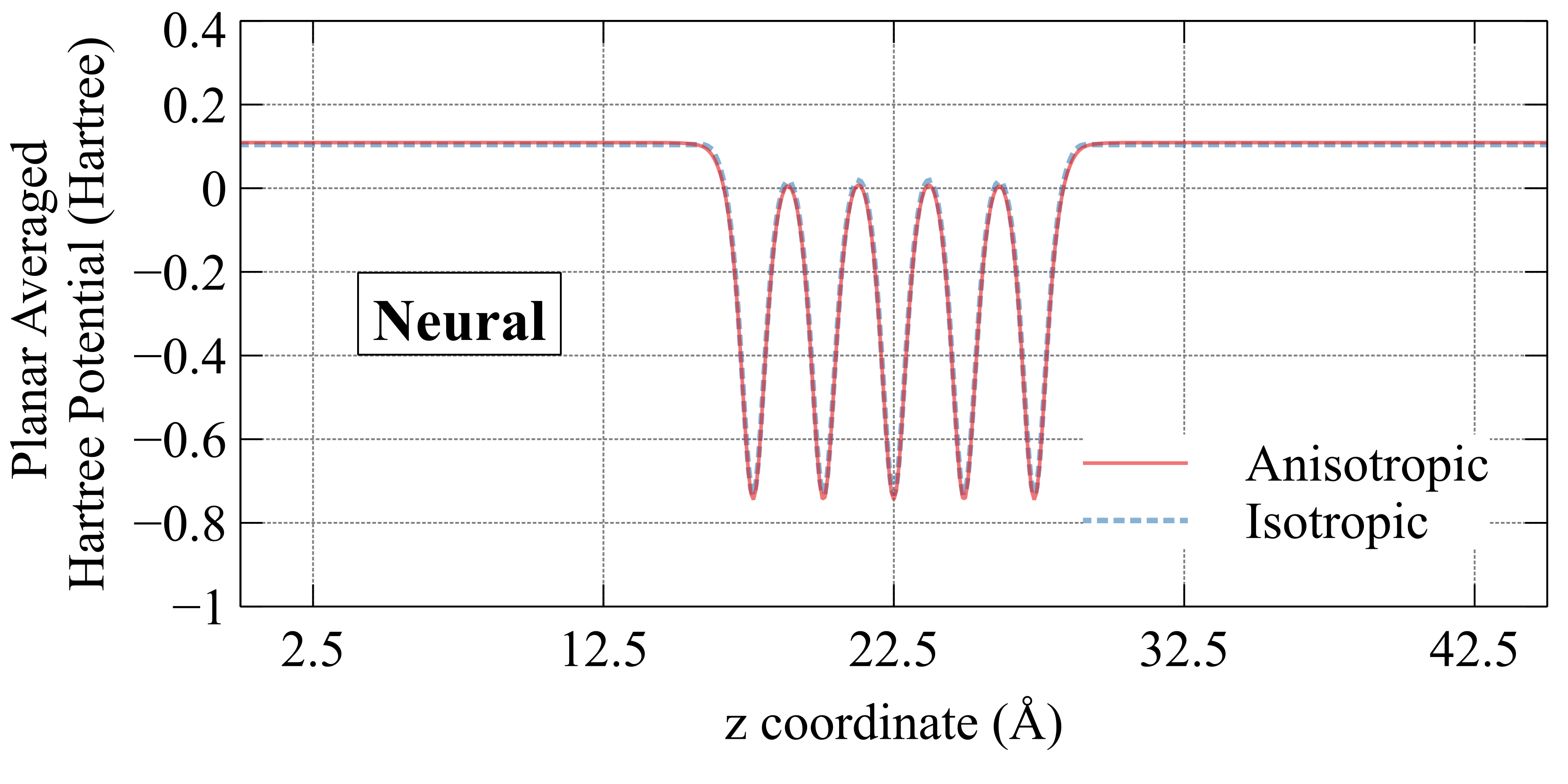}
\label{fig5b}
}
\subfigure[]{
\includegraphics[width=0.7\textwidth]{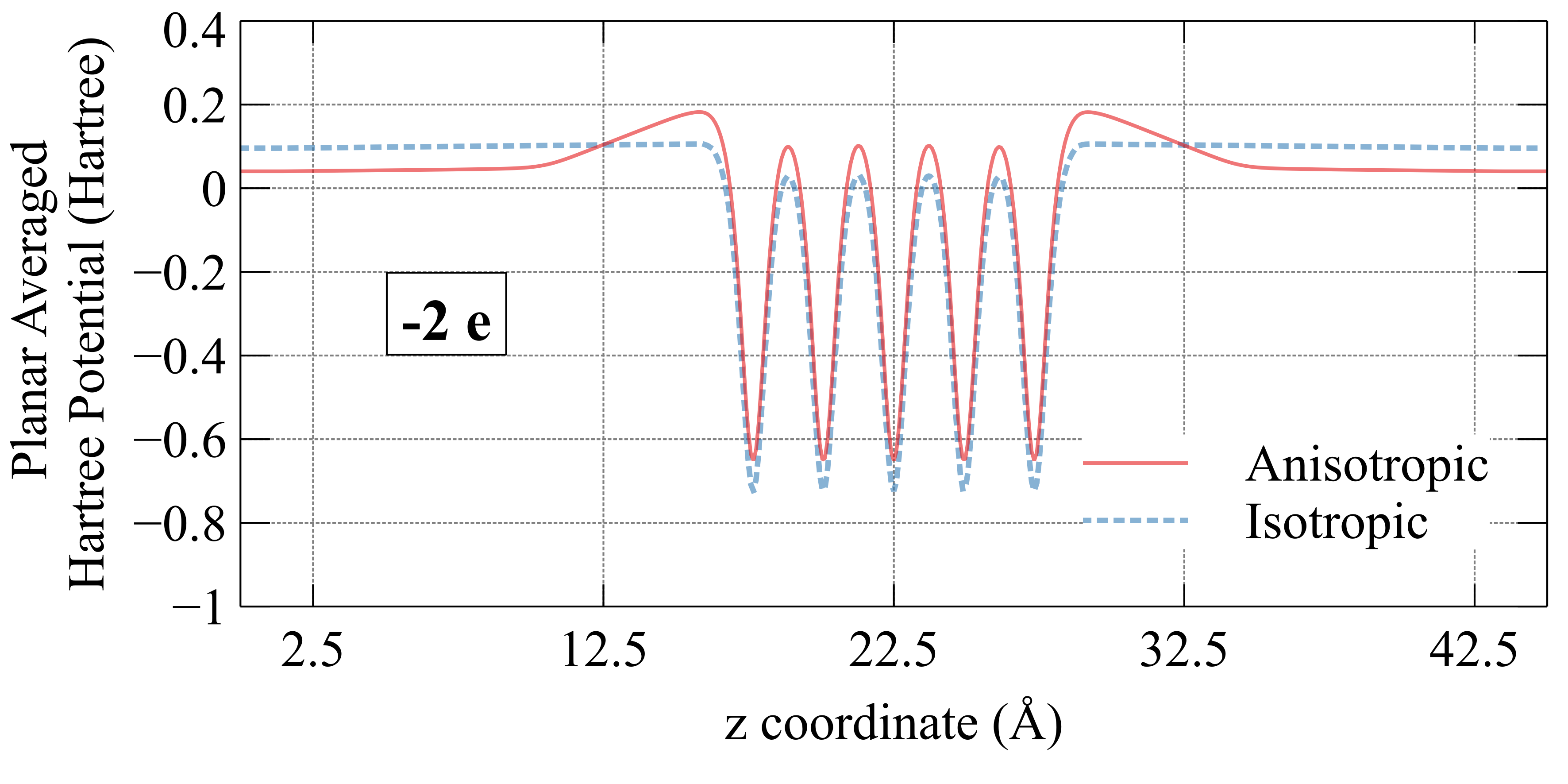}
\label{fig5c}
}
\caption{Profiles of the in-plane planar-averaged values of electrostatic potentials in the cases of +2 e (a), neutral (b), and -2 e (c) charge state as functions of z coordinates. The anisotropic cases were shown as solid red lines and the isotropic cases were shown as dotted blue lines.}\label{fig5}
\end{figure}

In the previous two sections, we have verified the reliability of FEAPS and the AICS model under vacuum condition, as well as in the case of isotropic dielectric tensor with no spatial variation within the $\rho_{\min}>\rho^{\text {el}}(\boldsymbol{r})$ regions. In this section, we then turn our attention to testing dielectric tensors that are anisotropic and exhibit spatial variation in the $\rho_{\min}>\rho^{\text {el}}(\boldsymbol{r})$ regions.\\
As reported previously in Tran et. al.’s study\cite{recent_anisotropic_dielectric_PRL_2023} which utilized a classical molecular dynamics approach in combination with an ac field method, the dielectric functions of interfacial water at the Ag(111)/water interface exhibited notably anisotropic and spatial variation along the surface normal direction (as shown in Fig. 3 in Ref. \citen{recent_anisotropic_dielectric_PRL_2023}). The in-plane dielectric constant is enhanced to around 92 from 78.3 when approaching the Ag(111) surface from water solvent along the surface normal direction. To mimic that (Fig. 3 in Ref. \citen{recent_anisotropic_dielectric_PRL_2023}), the following parameters in Eqs.~\eqref{eq9}--\eqref{eq11} were chosen for defining $\epsilon_{\text {dist,xx}}(\boldsymbol{r})$ and $\epsilon_{\text {dist,yy}}(\boldsymbol{r})$: $\epsilon_{\text {dist,xx}}^{\text {near}}=\epsilon_{\text {dist,yy}}^{\text {near}}=92.0$, $  d_\text {x}^0=d_\text {y}^0=5.0$ \AA, $ \alpha=0.5$ \AA, $  \beta=2.0$ ~\AA, and $\sigma=4.0$ ~\AA\ (the red curve in Fig.~\ref{fig3}). The out-of-plane dielectric constant is weakened from 78.3 to a much lower average value around 4 in the range of 0 \AA\ to 7.5 \AA\ above the outermost Ag atoms when approaching the Ag(111) surface. This average is obtained by applying a capacitor-in-series model to account for the near-surface vacuum gap and the negative dielectric function regions which will lead to numerical divergence and physical instability in solving Poisson equations. To mimic that, we used the parameters $\epsilon_{\text {dist,zz}}^{\text {near}}=2.0$, $ d_\text {z}^0=9.0$ \AA, $ \alpha=0.5$ \AA, $  \beta=2.0$ \AA, and $\sigma=4.0$ \AA\ (the blue curve in Fig.~\ref{fig3}), which resulted in an average dielectric constant around 1.7 incorporating the contributions from the vacuum gap and the transition region ($\rho_{\min}<\rho^{\text {el}}(\boldsymbol{r})<\rho_{\max}$). $\rho_{\max}=0.01$ and $\rho_{\min}=0.001$ were used as solute-solvent boundary parameters in all calculations in this section. Fig.~\ref{fig3} illustrates the in-plane and out-of-plane components of the dielectric tensor $\epsilon_{\text {dist,xx}}$ ($\epsilon_{\text {dist,yy}}$) and $\epsilon_{\text {dist,zz}}$, as a function of the distance $r_\text {z}$ from the surface.\\
The dielectric constants shown in Fig.~\ref{fig3} are almost invariant when $r_\text {z}$ approaches 0. This behavior was designed for ensuring a smooth transition at the $\rho_{\min}=\rho^{\text {el}}(\boldsymbol{r})$ boundary in practical calculations. Fig.~\ref{fig4} presents the in-plane planar-averaged values of the in-plane and out-of-plane dielectric functions, which were the combinations of the electron-density-derived ones (where $\rho_{\min}<\rho^{\text {el}}(\boldsymbol{r})$) and the three $\epsilon_{\text {dist,ii}}^{\text {near}}$ functions (where $\rho_{\min}>\rho^{\text {el}}(\boldsymbol{r})$), at a given z coordinate. The Ag(111) slab was inside the region where the z coordinate is larger than 17.5 \AA\ and small than 27.5 \AA. When leaving the surface from 17.5 \AA\ to 15 \AA\ and from 27.5 \AA\ to 30 \AA, the in-plane dielectric function increased from 1 to 92 and reached a plateau. When leaving the surface from 17.5 \AA\ to 11 \AA\ and from 27.5 \AA\ to 34 \AA, the out-of-plane dielectric function increased from 1 to 2 and reached a plateau. Afterwards, the dielectric functions varied with z coordinates according to the three $\epsilon_{\text {dist,ii}}^{\text {near}}$ functions as illustrated in Fig.~\ref{fig4}. Eventually, the dielectric functions all reached the dielectric constant of bulk solvent (in this case, 78.3 of water in room temperature). The values of the dielectric functions within the slab were manually set to 1 to eliminate unphysical implicit solvent, as also done in Sec.~4.1.\\
Given the form of the dielectric functions discussed and illustrated above, the DFT + AICS\&FEAPS SCF calculations for the Ag(111) slab with net charges of -2 e, 0 e, and +2 e were performed to obtain the resulting work functions and electrostatic (Hartree) potentials. For comparison, the corresponding isotropic calculation, in which $\epsilon_{\text {dist,xx}}^{\text {near}}$, $ \epsilon_{\text {dist,yy}}^{\text {near}}$, $ \epsilon_{\text {dist,zz}}^{\text {near}}$, and $\epsilon_{\text {bulk}}$ were all set to 78.3 to mimic the traditional SCCS calculation, was also performed by using FEAPS for each charge state. To neutralize the net charge of the Ag(111) slab, we added planar counter charge densities as described in Refs. \citen{pcc1,pcc2,our_gce_jctc_2024} placed on both sides of the slab, each located 0.8 \AA\ from the simulation‑cell boundary along the surface normal. Table~\ref{tab2} presents the calculated work functions (see the definition in Ref.\citen{our_gce_jctc_2024}) in each combination of anisotropic/isotropic dielectric models and charge states. The calculated work function for the anisotropic model increased monotonically from -0.52 eV, to 3.86 eV, and then to 8.78 eV as the net charge of the slab changed from -2 e, to 0 e, and then to +2 e. This behavior was also observed in the isotropic dielectric case (from 2.96 eV, to 3.44 eV, and then to 4.21 eV). This is consistent with expectations, since the electrons in the Ag(111) slab became more stable and the work function increased when more electrons were removed from the slab. What differs is that the work function varied more substantially in the anisotropic dielectric case. Note that our implicit-solvent reference lacks the vacuum–water surface potential (the computed ab initio values are in the range of 3 to 4 eV\cite{airwaterpot1,airwaterpot2,airwaterpot3}), and accounting for this offset removes the negativity of the work function -0.52 eV.\\
Fig.~\ref{fig5} presents the in-plane planar averages of the calculated electrostatic potentials as a function of z (along surface normal direction) coordinates and revealed the above discrepancy. We first check the results for the 0 e charge state (neutral slab) as shown in Fig.~\ref{fig5b}. It can be seen that the curves nearly overlap, except that in the anisotropic case the curve is slightly higher in the solvent region and slightly lower in the slab region compared to the isotropic case. The work function of the neutral slab is slightly higher in the anisotropic case (3.86 eV) than in the isotropic case (3.44 eV). This indicates that the dielectric‑continuum–induced reduction of the Ag(111) slab's vacuum work function is weaker in the anisotropic case (the calculated vacuum work function of the Ag(111) slab is 4.18 eV). This behavior is consistent with expectations, since in the dielectric plateau region near the slab in the anisotropic case, the in-plane dielectric constant increased from 78.3 to 92 whereas the out-of-plane dielectric constant decreased drastically from 78.3 to 2. It corresponds to a weaker dielectric screening and a more vacuum-like response. Due to electrical neutrality, both curves show no spatial variation along the z coordinate after reaching the $\rho_{\min}$ threshold. Fig. \ref{fig5a} presents the results for the positively charged slab (net charge +2 e). For the isotropic case, the averages became relatively stable (slowly increased) after electron density decreased below $\rho_{\min}$ away from the metal slab along the surface normal direction. In the anisotropic case, the electrostatic potential shifted downwards in the slab region and shifted upwards in the solvent body region, and the transition region between the two showed a steady increase away from the surface along surface normal direction. On the contrary, the results for the negatively charged slab (net charge -2 e) showed a reverse behavior, as shown in Fig.~\ref{fig5c}. For the isotropic model, the averaged potential again tended toward a plateau beyond $\rho_{\min}$ but with a slight decrease away from the surface along surface normal direction, whereas the anisotropic model shows an upward shift in the slab region, a downward shift in the solvent body, and a monotonic decrease across the interfacial transition.\\
For Ag(111) water interface we lack published plane-averaged electrostatic-potential benchmarks, but ab initio studies on Pt(111) water interface show that when electrons are transferred from water to the metal and the surface carries net negative charge, the plane-averaged potential, evaluated from inside the slab toward the electrolyte along the surface normal, switches from a steep rise within the metal to a pronounced downturn across the interfacial region (Fig. 1(a) in Ref. \citen{airwaterpot3}). It can be seen from Fig.~\ref{fig5c} (the solid red curve) that our model reproduced this spatial variation in the region 0 to $\sim$7 \AA\ away from the surface along the surface normal. However, isotropic, spatially uniform continuum models can only give a monotonic rise that saturates to an almost constant bulk value and cannot reproduce this interfacial profile (such as the dotted blue curve in Fig.~\ref{fig5c}). The oscillations near the first two hydration layers shown in Fig. 1(a) in Ref. \citen{airwaterpot3} likely arise from solvent layering, underscoring the need to refine the form of the surface-distance-dependent dielectric function $\epsilon_{\text{dist,ii}}(\boldsymbol{r})$ in future work.\\
\begin{table}[H]
    \centering
    \caption{Calculated work functions (in eV) of Ag(111) slab with -2 e, 0 e, and +2 e net charges, by using anisotropic and isotropic AICS\&FEAPS + DFT SCF. Planar counter charge densities were added to completely screen the net charge of the Ag(111) slab.}
    \label{tab2}
\begin{tabular}{| l | l | l |}
\hline
\multicolumn{3}{|l|}{Work function (eV)} \\
\hline
  & Anisotropic & Isotropic \\
\hline
+2 e charge & 8.78 & 4.21 \\
\hline
0 e charge & 3.86 & 3.44 \\
\hline
-2 e charge & -0.52 & 2.96 \\
\hline

\end{tabular}

\end{table}
\subsection{4.3 OH* Adsorption on Ag(111) Surface}
In Secs. 4.1 and 4.2, we validated the consistency between the energies and analytical forces, and tested the computed electrostatic potentials. In this section, we turn to the adsorption of OH* on the Ag(111) surface, a phenomenon commonly encountered in various physicochemical processes at the Ag(111)/water interface\cite{jacs_oh_ag111}. We employed the AICS method based on FEAPS to optimize the adsorption geometry of OH* on the Ag(111) surface. For comparison, we also optimized the adsorption geometries both in the vacuum enviroment and in the isotropic dielectric water enviroment by employing the isotropic SCCS method. In the calculation using the AICS method, the parameters (in Eqs.~\eqref{eq9}--\eqref{eq11}) used in Sec. 4.2.3 were chosen for defining $\epsilon_{d i s t, x x}(\boldsymbol{r})$, $\epsilon_{d i s t, y y}(\boldsymbol{r})$, and $\epsilon_{d i s t, z z}(\boldsymbol{r})$, except that $d_\text {x}^0$, $ d_\text {y}^0$, and $d_\text {z}^0$ were all adjusted to 10.0 \AA\ to ensure a smooth transition of the dielectric constant across the OH*/implicit‑solvent boundary. $\rho_{\max}=0.001$ and $\rho_{\min}=0.0001$ were used as solute-solvent boundary parameters in all calculations in this section. RPBE\cite{rpbe} exchange correlation functional together with semi-empirical dispersion corrections D3\cite{D3,vdw1,vdw2}, which were shown to correctly reproduce wetting behavior of water on closed-packed metal surfaces\cite{why_rpbe_1, why_rpbe_2}, were used in all calculations in this section. MOLOPT basis sets\cite{molopt_basis} were used for the H element, DZVP-MOLOPT-SR-GTH-q1, and the O element, DZVP-MOLOPT-SR-GTH-q6. The norm-conserving Goedecker, Teter, and Hutter pseudopotentials\cite{gth_pp1,gth_pp2,gth_pp3} were used for H element, GTH-PBE-q1, and the O element, GTH-PBE-q6. All other computational settings followed those in Sec. 4.2.3.\\
The geometry optimizations were performed in a sequential manner for a Ag(111) slab + 2OH* model in which the two OH* groups adsorbed on the two (111) surface symmetrically. The geometry was first optimized in the vacuum environment. It was further optimized by using SCCS and then by using AICS + FEAPS. The top and side views of the optimized geometries are shown in Fig.~\ref{figs3}. Table~\ref{tab3} presents the Ag-O and O-H bond lengths and the angles between the O-H bond and the Ag(111) surface, for the optimized geometries. Compared to vacuum (2.26 \AA), implicit solvent led to more stretched Ag–O bond lengths (SCCS: 2.28 \AA, AICS: 2.31 \AA). Under our dielectric function settings, the AICS model resulted in a larger stretching of the Ag–O bond length than the SCCS model. Compared to the Ag-O bond length, the O–H bond length was less affected by implicit solvent. 
The angle between the O-H bond and the Ag(111) surface in vacuum (38\textdegree) was slightly suppressed (to 34 \textdegree) when the isotropic implicit solvent was imposed (by using SCCS). The angle was greatly reduced to 17 \textdegree when our anisotropic implicit solvent was imposed (by using AICS). The reason is that the in-plane dielectric function surrounding the OH* adsorbate was 92, which resulted in a high degree of screening and in-plane stability of the dipole moment of OH*. The out-of-plane dielectric function surrounding is only 2, which resulted in a very minor screening and stability of the dipole moment in the out-of-plane direction. The OH* group tended to rotate towards the plane parallel to the surface to lower the energy. Although there is no direct evidence for the O–H bond angles of adsorbed OH* on Ag(111) in aqueous environments due to a lack of computational and experimental studies, a previous ab initio molecular dynamics study by Partanen and Laasonen\cite{pccp_2024_oh_angle}, which modeled the interface with an explicit water slab, found that these angles on Pt(111) mainly fell within the range of 15 \textdegree to 25 \textdegree.\\
\begin{table}[H]
    \centering
    \caption{Ag-O and O-H bond lengths and angles between the O-H bond and the Ag(111) surface of the optimized geometries.}
    \label{tab3}
\begin{tabular}{lccc}
\hline
\textbf{} & \multicolumn{2}{c}{\textbf{Bond length (\AA)}} & \textbf{O-H/surface angle (°)}\\
          & Ag--O & O--H & \\
\hline
Vacuum    & 2.26  & 0.97 & 38 \\
SCCS      & 2.28  & 0.97 & 34 \\
AICS      & 2.31  & 0.98 & 17 \\
\hline
\end{tabular}
\end{table}
\subsection{4.4 Parallel Performance}
\begin{figure}[H]
\centering
\includegraphics[width=\textwidth]{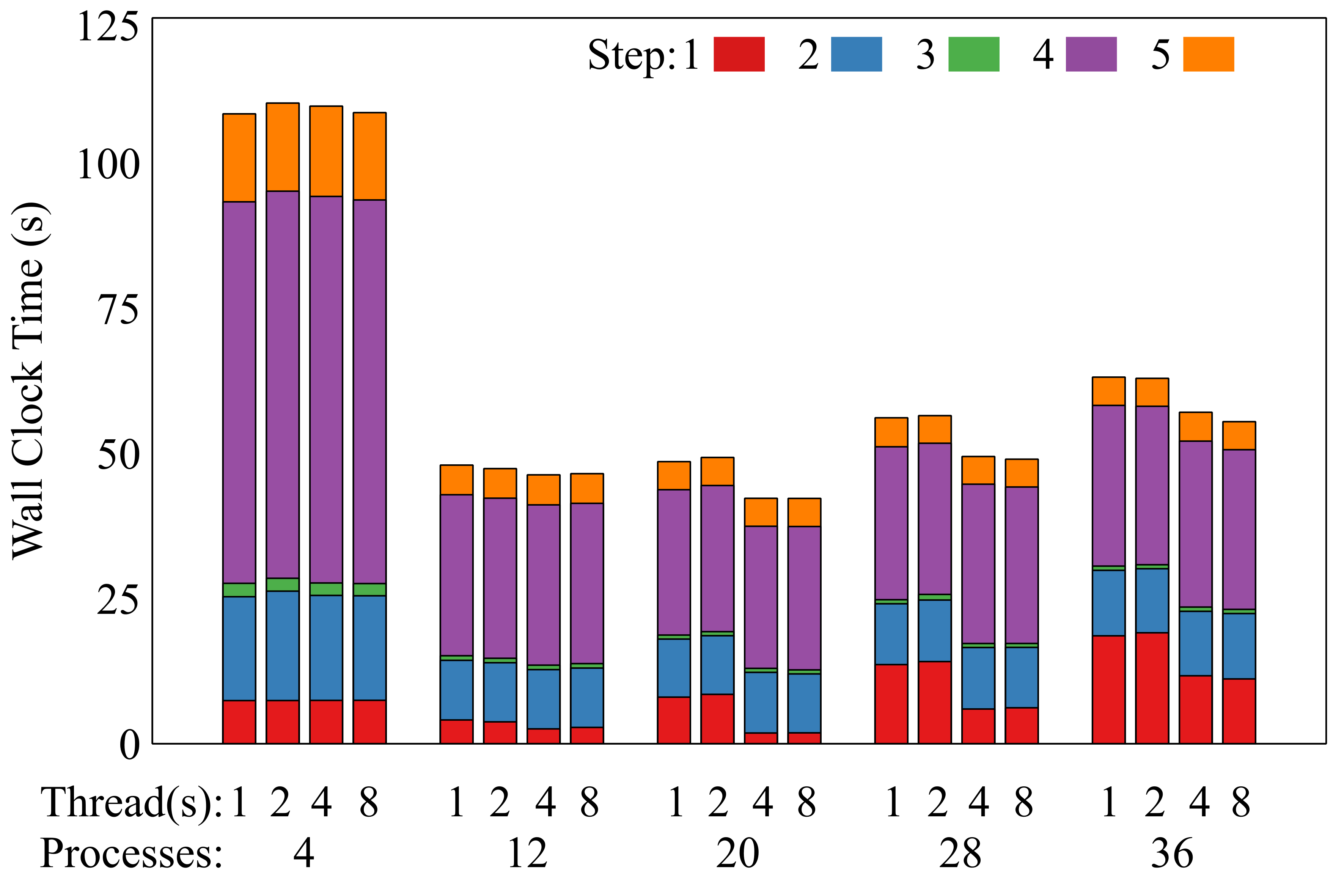}
\caption{Wall clock times (stacked bars, in seconds) of steps 1 to 5 during the complete executions of the FE Python module from initialization to solution. The tests were performed over a range of hybrid MPI/OpenMP configurations, varying both the number of MPI processes and the OpenMP threads per process.}\label{fig6}
\end{figure}

\begin{figure}[H]
\centering
\includegraphics[width=\textwidth]{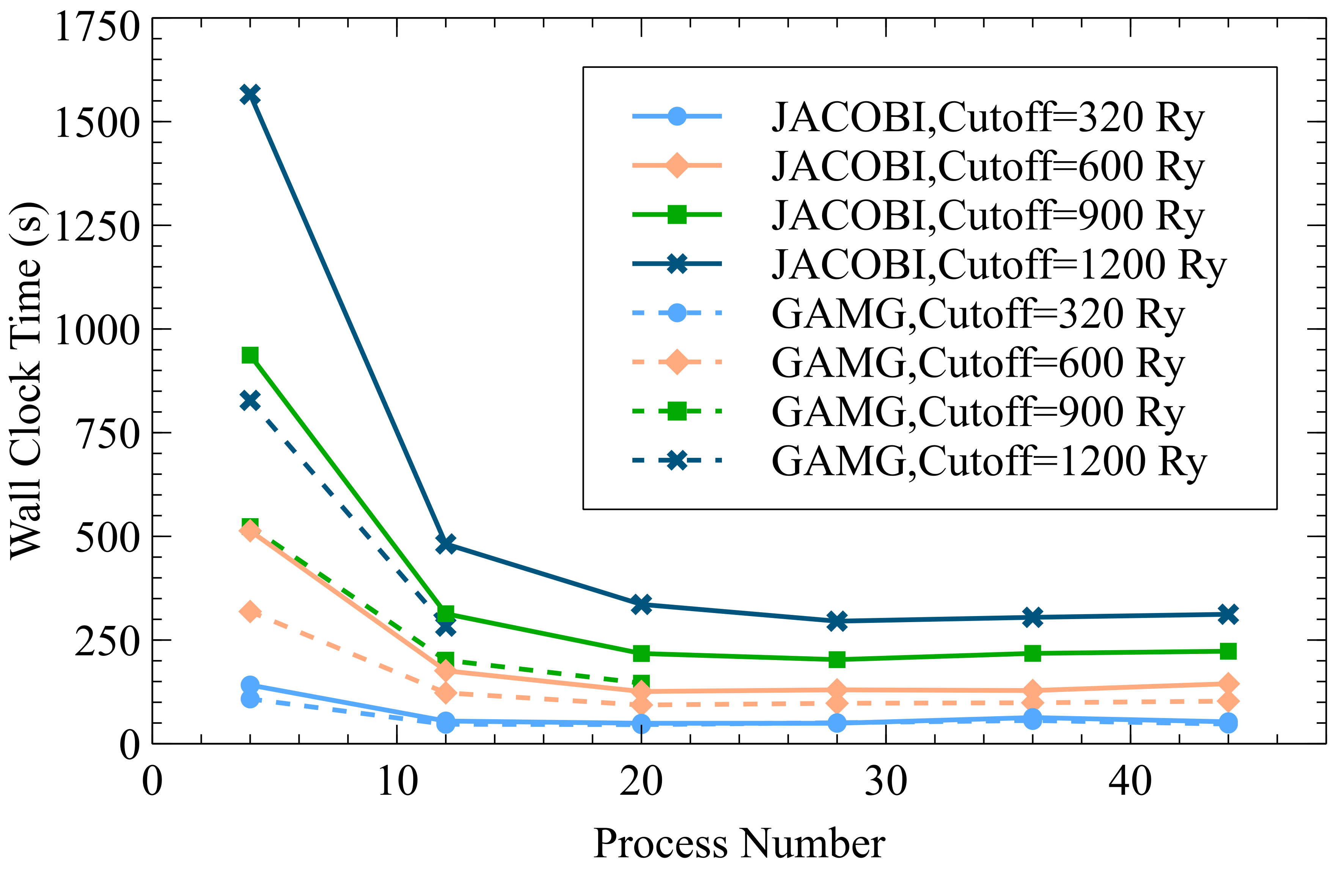}
\caption{Wall clock times of the complete FAEPS executions under various combinations of the number of MPI processes, preconditioner, and cutoff energy.}\label{fig7}
\end{figure}

Finally, we evaluated the parallel performance of the implemented FEAPS using the Ag(111)/implicit water interface model. The slab model was charge-neutral and embedded in an anisotropic dielectric continuum. The computational parameters were the same as the ones used in the case of the anisotropic water and charge neutral slab in Sec. 4.2. Each compute node was equipped with four 72-core NVIDIA Grace-Hopper ARM sockets, providing a total of 288 CPU cores and up to 512 GB of memory per node. A series of single-iteration SCF calculations were performed, during which the complete workflows of the finite-element anisotropic Poisson solver—from initialization to solution—were executed. As the data exchange between the main Fortran code of CP2K and the Python-based finite element (FE) solver was implemented via direct memory address passing, communication overhead is negligible. Accordingly, we report only the wall clock times associated with the FE solver executions. In a complete SCF calculation, the FE mesh and Poisson solver are initialized once at the beginning of the SCF cycle. In the subsequent SCF steps, only data on the existing mesh are updated, and the solver proceeds without reinitialization. All calculations were conducted on a single compute node. A range of parallel configurations was evaluated by varying the number of MPI processes (4, 12, 20, 28, and 36) and the number of OpenMP threads per process (1, 2, 4, and 8).\\
Fig.~\ref{fig6} shows the wall-clock times (in seconds) as stacked bars during the complete executions of the FE Python module from initialization to solution, at the SCF step where the AICS model and FEAPS were activated for the first time. An execution was divided into five stages (step 1-5, represented in five different colors). Step 1 built the geometry mesh which was the same size as the one used in CP2K DFT calculations to represent real-space functions. When building the mesh, a 3-dimensional box was uniformly divided into small voxels in the same way as the simulation box was divided in CP2K. Step 2 was to redistribute the mesh among the processes in the way described in Sec. 4 and create the FE function space. We note in passing that Step 1 and 2 (FE mesh initialization) were executed only once at the beginning of the whole process of the SCF iterations. Step 3 was to interpolate dielectric functions and solute’s charge density data to the corresponding FE functions. Step 4 assembled the linear problem (solver initialization, only be executed once during SCF iterations) and solved the equation. Step 5 was to transfer the electrostatic potential solution to the real-space grid function in CP2K.\\
One can find from Fig.~\ref{fig6} that step 2 and step 4 are the most time-consuming ones when the number of processes is less than or equal to 20. This indicates that avoiding unnecessary reinitialization in Steps 2 and 4, and instead updating data only when necessary, is important for improving computational efficiency. The computation time of each step decreases as the number of processes increases from 4 to 12. With further increases in the number of processes, the computation time remains nearly unchanged. The computation time of Step 1, which involved calling DOLFINx’s built-in function "dolfinx.mesh.create\_box", mostly increases when the number of processes increases from 12 to 36. This might imply that the algorithm limit of the function was reached, or the implementation of the function can still be improved in the future. In addition, in the cases of process number 12, 20, 28, and 36, a small amount of reduction of computational time is observed when the thread number increases from 2 to 4.\\
Fig.~\ref{fig7} shows the total execution time of FEAPS as a function of the number of MPI processes (OpenMP thread number was 1). Each curve corresponds to a specific combination of preconditioner ("GAMG" or "JACOBI") and cutoff energy (320 Ry, 600 Ry, 900 Ry, or 1200 Ry), where the cutoff determined the resolution of a real-space on the CP2K grid and directly affected the number of voxels. In each case of the combinations, the wall clock time decreased drastically when the number of used processes increased from 4 to 12 and became stable when the number of processes was over 20. At each cutoff, the JACOBI preconditioner yielded longer wall-clock times than GAMG. We conclude that to save computational time, the GAMG preconditioner is a better choice. However, it should be noted that the GAMG preconditioner has a higher memory demand. In the case of the GAMG preconditioner + 900 Ry cutoff and the GAMG preconditioner + 1200 Ry cutoff, several data points are missing due to memory exhaustion (512 GB limit).\\
\section{5 Conclusion}
In this work, we developed an anisotropic interface continuum solvation (AICS) method, which incorporates anisotropy and spatial variation of dielectric tensors along the surface normal direction, to accurately simulate the anisotropic dielectric behaviors of solvent liquids near solid-liquid interfaces. We implemented the AICS method including the derived analytical expressions for the electrostatic contributions to the Kohn–Sham potentials and the atomic forces within the CP2K software package. To solve the anisotropic Poisson equations with anisotropic dielectric tensors, we developed a parallel finite-element anisotropic Poisson solver (FEAPS) based on the FEniCSx computational platform and its interface with CP2K. For the Ag(111) surface slab model, rigorous numerical validations showed excellent agreement of analytical atomic forces with finite-difference results, and the electrostatic potentials computed under vacuum and isotropic solvent setups matched closely with conventional FFT-based vacuum DFT and DFT + SCCS calculations, respectively. In the anisotropic solvent (water) environment characterized by enhanced in-plane and reduced out-of-plane dielectric functions near the charged Ag(111) interface, our method demonstrated notable differences in computed electrostatic potentials and work functions compared to isotropic models. Furthermore, geometry optimizations of OH* adsorption on Ag(111) revealed that the anisotropic solvent environment greatly influenced the orientation of the adsorbed OH*, highlighting the necessity of incorporating dielectric anisotropy into continuum solvation models for accurate modeling of interfacial phenomena of a solid-liquid interface. The finite-element anisotropic Poisson solver demonstrated good parallel scalability, with significant reductions in wall-clock time observed up to 12 MPI processes. The GAMG preconditioner offers better performance than JACOBI, although at the cost of increased memory usage. Overall, the proposed method and its software implementation offer a reliable and efficient framework for first-principles modeling of realistic solid–liquid interfaces, where the interfacial liquid exhibits anisotropic dielectric behavior driven by interfacial effects.
\begin{acknowledgement}

This work was supported by the University of Zurich and SNSF Sinergia Project CRSII5\_202225. This work was supported by the grants from the Swiss National Supercomputing Centre (CSCS) under project ID s1277 and lp11. We gratefully acknowledge Andrey Sinyavskiy for his helpful discussions and help with Python programming language, the installation of FEniCSx on the Alps Daint supercomputer.

\end{acknowledgement}

\bibliography{achemso-demo}
\end{document}